\documentclass[review]{elsarticle}

\usepackage{amsmath,amssymb,amsfonts}
\usepackage{textcomp}
\usepackage{hyperref}
\usepackage{amsmath}
\usepackage{amssymb}
\usepackage{graphicx}
\usepackage{float}
\usepackage{algorithm}
\usepackage{algpseudocode}
\usepackage{comment}
\usepackage{subcaption}

\begin{document}
\begin{frontmatter}
\title{Deep Learning Based Dynamics Identification and Linearization of Orbital Problems using Koopman Theory}

\author{George Nehma}

\author{Madhur Tiwari}

\author{Manasvi Lingam}

\affiliation{organization={Department of Aerospace, Physics and Space Sciences, Florida Institute of Technology},
            addressline={150 W University Blvd}, 
            city={Melbourne},
            postcode={32901}, 
            state={FL},
            country={USA}}

\clearpage
\begin{abstract}
The study of the Two-Body and Circular Restricted Three-Body Problems in the field of aerospace engineering and sciences is deeply important because they help describe the motion of both celestial and artificial satellites. With the growing demand for satellites and satellite formation flying, fast and efficient control of these systems is becoming ever more important. Global linearization of these systems allows engineers to employ methods of control in order to achieve these desired results. We propose a data-driven framework for simultaneous system identification and global linearization of the Circular, Elliptical and Perturbed Two-Body Problem as well as the Circular Restricted Three-Body Problem around the L1 Lagrange point via deep learning-based Koopman Theory, i.e., a framework that can identify the underlying dynamics and globally linearize it into a linear time-invariant (LTI) system. The linear Koopman operator is discovered through purely data-driven training of a Deep Neural Network with a custom architecture. This paper displays the ability of the Koopman operator to generalize to various other Two-Body systems without the need for retraining. We also demonstrate the capability of the same architecture to be utilized to accurately learn a Koopman operator that approximates the Circular Restricted Three-Body Problem.
\end{abstract}
\end{frontmatter}

\section{INTRODUCTION}
The rapid growth in motivation to explore space in the last fifteen years has sparked a number of significant missions to planets such as Mars and Venus, moons such as Enceladus and Titan, as well as our own Moon. From interplanetary missions to the placement of satellites within our own planet's orbit, we would not be able to achieve the goals of these endeavors without understanding and exploiting the natural dynamics of these systems. Much effort has been spent in determining how exactly the dynamics of bodies, either natural or man-made, are influenced, and how engineers can develop guidance, navigation, and control systems to complete these extraordinary tasks. Today, thousands of satellites orbit around our Earth in a variety of different orbits, and the governing equations of motion for all of these satellites are the equations originating from the N-Body Problem. In the N-Body equations of motion, the only force that imparts energy onto the bodies is gravity \cite{SJA03}, and dependent on the complexity of the solution required and the nature of the system in question, the number of bodies varies. Most often in celestial dynamics, the Two-Body Problem (2BP) and Circular Restricted Three-Body Problem (CR3BP) are analyzed and used \cite{2bp1,2bp2,2bp3,VS67,BG97,CM90,MCG98,MD99,MV05,VAK16}. As we endeavor to reach the Moon again, and move forth to Mars, the analysis and use of the CR3BP will become even more prevalent \cite{CM90,MV05,VAK16,DARVISH201512,DU2024109048}.

In the case of artificial satellites orbiting Earth, they are all governed by the 2BP with perturbations \cite{satellites,KFW15}. With the growing number of these satellites being put into orbit, as seen with the explosion of SpaceX Starlink satellites, the need for a fast, efficient and guaranteed method for propagating and controlling the complex dynamics of these bodies is becoming more evident. The linearization of the 2BP has been achieved through the Clohessy-Wiltshire relative motion equations \cite{chweqns}, under extensive assumptions, but the discovery of a globally linear representation of this problem has yet to be uncovered.

The case of the CR3BP is a well-known and studied reduction of the Three-Body problem, to which no global analytical solutions exist \cite{CM90,MV05,VAK16}. Instead, quasi-periodic, periodic and hyperbolic invariant manifolds present themselves as special solutions, which can inform us on the dynamics of the CR3BP \cite{VS67,CR3BP,CM90,MV05,VAK16}. This system is of importance because the Earth-Moon system can be treated as such \cite{MCG98}, with the orbit of the Moon around the Earth having an eccentricity of 0.0549 \cite{e_Moon}. The Earth-Moon system has five equilibrium points, the so-called Lagrange points \cite{MD99}, around which it is possible to generate periodic orbits that are integrable. Much like the 2BP, efficient, accurate linear representation of the CR3BP would further advance the exploration of space, by reducing the complexity of creating control systems capable of being used in these systems. 

The control of nonlinear systems has long been a challenge for engineers. Linear control systems often have straightforward and useful methods in order to achieve theoretical guarantees in the controllability, observability, and stability of the system. Current techniques provide some relief to this problem, however the horizon upto which these techniques can be relied on is measurably short. This leads to frequent linearization calculations around specific operating points, increasing computational cost and time when performing control maneuvers and state estimation. A globally linear representation of the dynamics of the system would be an ideal solution, eliminating the drawbacks of traditional linearization techniques. Unfortunately, finding such global linearizations is still a prominent issue, but with the recent advancements in Koopman theory, the prowess of machine learning and advancements in computational power, this ideal solution may not be as unattainable as once thought. 

Koopman theory, first proposed in 1931 \cite{Koopman1931}, has gained traction over the last few years as a solution for finding global linearization of nonlinear systems. In a nutshell, the theory states that the dynamics of a nonlinear system can be described linearly by an infinite-dimensional Koopman operator. Due to its infinite dimension, for practical use, it is typically approximated using data-driven methods such as Extended Dynamic Mode Decomposition (EDMD) \cite{EDMD}. The process of analytically implementing these approximating methods such as EDMD or Hankel-DMD \cite{Arbabi_2017} on the nonlinear system to find an accurate approximation is not only highly involved, but requires an innate understanding of the method in order to select a reasonable, diverse and encompassing dictionary of basis functions. These basis functions are the building blocks from which functions known as observable functions are constructed. These observable functions are then what is applied to the data in order to lift dynamics to the appropriate dimension. However, the main focus of our work is in the construction of a feed-forward neural network (NN) that implements EDMD, making it easier to find an approximate general Koopman operator linearization applicable to a region of state space, that is time invariant and constant. Because of the high interest in Koopman Theory as a method for globally linearizing the nonlinear dynamics of a problem, a great deal of research has been conducted in a wide variety of fields. Applications in fluid dynamics \cite{constanteamores2024datadrivenkoopmanoperatorpredictions}, robotics \cite{koopmanrobot} and control systems \cite{koopmancontrol} have all been investigated. Whilst the utilization of Koopman in order to develop linear systems capable of control has been researched heavily, with many models including optimal controls \cite{jin2023datadrivenoptimalcontroltethered}, Model Predictive Controllers (MPC) \cite{MAO2024109515}, Linear Quadratic Controllers (LQR) \cite{manaa2024koopmanlqrcontrollerquadrotoruavs}, all being successfully developed through the implementation of EDMD with control (EDMDc) \cite{edmdc} via analytical solutions or the development of NN's. Another advantage of using a data-driven method is that it does not require system knowledge, and the nonlinear system can be completely unknown \cite{brunton2021modern,kaiser2021data}.

Neural Networks, although having a history that spans almost 40 years, have found a resurgence in the last 10-15 years, propelled by the work of LeCun et al. \cite{lecun2015deep}, who were able to analyze the ImageNet dataset and accurately categorize image classes within the dataset. Since then, the advancements of big data coupled with computational power has allowed Deep Neural Networks (DNNs) to grow rapidly, and be integrated into many fields of research and development. The power of DNN's lies in their ability to represent any arbitrary function, including the Koopman observable functions required to linearize the nonlinear dynamics \cite{Lusch_2018}. This is possible through a DNN with an adequate number of hidden neurons and a linear output layer, satisfying the universal approximation theorem \cite{cybenko1989approximation,hornik1990universal,hornik1989multilayer}. The use of a DNN in the approximation of the Koopman operator relieves us from the need to determine a satisfactory set of basis functions to which the Koopman eigenfucntions are mapped from, which is often difficult and can lead to poor approximations if not chosen correctly. 

The linearization of orbital (two and three-body) dynamics is an application where this global linearization can be highly beneficial for mission engineering. By reducing the computational cost, already limited CPU power can be diverted to more demanding tasks. In particular, a Koopman framework could be used in order to perform the state estimation tasks required for close-proximity maneuvers such as cluster formation, orbital changes, station keeping and rendezvous. In recent years, the rising abundance of satellite clusters associated with projects such as SpaceX's Starlink network and LISA (Laser Interferometer Space Antenna) \cite{LISA,AAA23} have revived the need for a fast, efficient and guaranteed control of satellites relative to one another.

Current state of the art of the linearization of the 2BP and CR3BP includes the use of Koopman Operator Theory. The relative, Clohessy-Wiltshire equations were successfully modeled by Koopman Theory to relieve some of the underlying assumptions as shown by \cite{servadio2022koopmanoperatorcontroloptimizationrelative}. Other models, including an oblate planet problem \cite{Koopmanoblate}, Sun-synchronous orbit \cite{Arnas_2021} and Earth-Moon L1 Lagrange point system \cite{servadio2021dynamicsnearthreebodylibration} have all been successfully linearized with the Koopman operator. On the other hand, \cite{cr3bpkoopman,2bp1, servadio2024uncertaintypropagationfilteringkoopman} were able to utilize the Koopman operator to study and analyze the separate challenges related to modeling the CR3BP, including models around co-linear Lagrange points and uncertainties in the propagation of the dynamics. The work \cite{cr3bpedmd} however, implements EDMD to analyze the CR3BP in regions outside of these Lagrange points, expanding on the breadth of knowledge base of Koopman applications on orbital problems. Because of the numerical complexity of both the 2BP and CR3BP, the treatment of the theory in this domain is often contextualized to specific conditions rather than generalized to the entire problem as a whole.

Numerous advancements in the framework are similar to the one presented in this paper. Not only does the model generate a globally linear, time-invariant representation of the nonlinear dynamics, removing the need for applications of system identification, but we can also use this model's prediction capability to work as a state estimator, eliminating the need for a Kalman Filter or its variants. The now globally linear system also no longer requires the burden of frequent linearizations such as Taylor Series expansions. In the case of the 2BP, assumption-heavy and limiting equations such as the Clohessy-Wiltshire equations may now be discarded in favor of this linear system. 

This work extends the framework developed in Tiwari et al. \cite{Tiwari}. The framework of the DNN developed by Tiwari et al. is kept the same, with improvements added for this application, being the modification of the loss function to be more suited to the significantly more complex and nonlinear dynamics that is the 2BP and CR3BP. We emphasize the significance of the CR3BP dynamics and the simplicity of our framework that is able to provide a global linearization around a L1 point. Although the previous application included control, it is not incorporated for the purposes of this paper when the methodology is applied to both the 2BP and CR3BP. To the best of our knowledge, this is the first time that both the 2BP and the CR3BP have been globally linearized with reasonable precision. 

The contributions of this paper are twofold:
\begin{itemize}
    \item First we demonstrate the ability of the learned Koopman operator to globally linearize both the Circular and Elliptical Two-Body problem centered around the Earth for orbits with an apogee altitude ranging between 200-30,000 km. We also demonstrate the ability of the same network, without additional training, to generalize to other two body systems, specifically around the Moon and Jupiter. We demonstrate that the approximation is accurate and that various invariant properties of the circular orbit are conserved. We also demonstrate that the prediction is reasonably accurate for significantly extended time horizon predictions, namely for 10 orbital periods. And we also show the capability of our model to capture non-periodic perturbations such as J2 and Solar Radiation Pressure (SRP).
    \item  Second, we demonstrate the applicability of the Koopman operator to the more complicated Circular-Restricted Three-Body Problem (CR3BP). We globally linearize the periodic orbit of a satellite that orbits the Earth, including the influence of the Moon, with the initial position of the satellite being close to the L1 Lagrange point. We also analyze the accuracy of the approximation by studying the conservation of the Jacobi constant in the CR3BP, showing that our model can adequately capture the value of this constant throughout the evolution of the dynamics. 
   
\end{itemize}

\section{NONLINEAR DYNAMICS AND THE KOOPMAN OPERATOR}

\subsection{Koopman Operator Theory}
\subsubsection{Koopman Operator}
The Koopman operator theory states that a nonlinear dynamical system can be transformed into an infinite-dimensional linear system \cite{Koopman1931}. Consider a dynamical system; the nonlinear, continuous dynamics are propagated by the following mathematical expression:

\begin{equation}
    \frac{d}{dt}\boldsymbol{x}(t) = \boldsymbol{F}(\boldsymbol{x}(t)) 
\end{equation}

where,  \(\boldsymbol{x}(t) \in \mathbb{R}^n \) is the system state at time \(t\) and \(\boldsymbol{F}\) is the function that describes the evolution of the state in the continuous sense. This continuous system can also be modeled in a discrete-time representation by evaluating the solution to the system at finite, discrete time intervals \(\Delta t\), such that \(\boldsymbol{x}_k = x(k\Delta t).\) Hence, the discrete-time dynamics are represented as:

\begin{equation} \label{eqn2}
    \boldsymbol{x}_{k+1} = \boldsymbol{f}(\boldsymbol{x}_k)
\end{equation}

where \(\boldsymbol{x}_k \in \mathbb{R}^n \) is the system state, \(k\) is the current time step, and \(\boldsymbol{f}\) is the function that evolves the system states through state space. We can then define observables, which are real-valued functions of the system state: \(g : \mathbb{R}^n \rightarrow \mathbb{R}\) such as: \([x^2, x^3, sin(x), cos(x)]\) for instance. The continuous time Koopman operator, \(\mathcal{K}\) and discrete time operator \(\mathcal{K}_{\Delta t}\), is therefore defined such that, for any observable function $g$,

\begin{equation} \label{eqn3}
    \mathcal{K}g = g\circ \boldsymbol{F},
\end{equation}

\begin{equation} \label{eqn4}
    \mathcal{K}_{\Delta t}g(\boldsymbol{x}_k) = g(\boldsymbol{f}(\boldsymbol{x}_k(t)))
\end{equation}

where \(\circ\) is the composition operator (explicitly defined in \ref{eqn6}). We can now apply this operator to the continuous and discrete-time system defined previously to arrive at:

\begin{equation} \label{eqn5}
    \mathcal{K}g = g\circ \boldsymbol{F}(\boldsymbol{x}(t)) = \frac{d}{dt}g
\end{equation}

\begin{equation} \label{eqn6}
    \mathcal{K}_{\Delta t}g(\boldsymbol{x}_k) = g(\boldsymbol{f}(\boldsymbol{x}_k)) = g(\boldsymbol{x}_{k+1}).
\end{equation}

It is evident in Equation \ref{eqn6},  that the Koopman operator \(\mathcal{K}_{\Delta t}\), propagates the observable function of a state \(g(\boldsymbol{x}_k)\) through time to the next time step. An important note is that for this work, we only require use of the discrete-time dynamics, and all further reference to \(\mathcal{K}\) is in the discrete-time sense \cite{snyderkoopman}. However, an impractical limitation of this operator is that it is formed in an infinite-dimensional space, therefore, numerical methods are required in order to approximate the operator in a finite dimensional space so that we may be able to access its utility in the control and evolution of dynamic systems. 

\subsubsection{Dynamic Mode Decomposition and its Extensions}

Determining the set of observable functions that span a Koopman invariant subspace, required to lift the states, is not a trivial task. Hence, a number of methods have been derived such as Dynamic Mode Decomposition (DMD), which aims to extract the dynamic modes of a given set of data. These modes can be interpreted as the generalization of the global stability modes and therefore project the underlying physical mechanisms of the system \cite{DMD}.

DMD has also been extended into Extended Dynamic Mode Decomposition (EDMD), allowing better approximations of the Koopman eigenvalues and eigenfunctions of the system, hence a closer representation of the nonlinear system \cite{EDMD,klus2016towards}. Essentially DMD is an approximation using monomial basis functions, analogous to a first-order Taylor expansion, whilst EDMD retains a higher number of terms in the expansion, thus allowing for a better approximation of a wide array of problems. 

In all applications of DMD, it is worth noting that the observable functions are built from a set of basis functions. These basis functions which can be any set of functions, such as monomials, radial basis functions (RBF's) or any other combination, are often chosen by the user. There is a strong connection between the basis functions and DMD's ability to learn the eigenfunctions and eigenvalues needed to create a more accurate representation of the Koopman operator \cite{kaiser2021data}. Hence, it is largely critical that the determination of the basis functions be made correctly. 

\subsubsection{Koopman Approximation Algorithm}

In this work, the finite dimension Koopman operator is approximated using the combination of EDMD with deep neural networks (DNNs). The procedure in which this is achieved is based off of our method in our previous work, \cite{Tiwari}, with the work presented herein being an extension of the architecture and structure presented previously with physics-informed modifications to the loss function, in order to better suit the problem. The DNN is used in order to learn and define the set of observables that are used to lift the states, contrary to other methods in which these observable functions are selected manually as a dictionary \cite{KoopmanInputControl,folkestad2020extended,yawcontrol}. This enables us to save significant time by not requiring the determination of a dictionary of basis functions, which need to be hand crafted for every problem, that serve to create the observable functions. Subsequently, the finite-dimensional approximation to the Koopman operator is calculated by utilizing least-squares regression (see Algorithm \hyperref[alg:1]{1}) \cite{Tiwari}.

\begin{figure*}[ht!]
    \centering
    \includegraphics[width=\columnwidth]{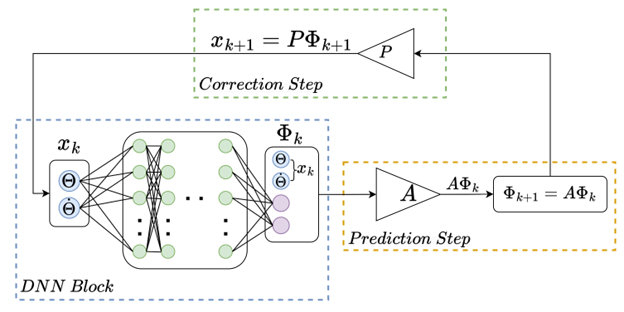}
    \caption{Deep learning framework including the prediction and correction steps. \cite{Tiwari}}
    \label{fig:net}
\end{figure*}

\begin{algorithm}
\caption{Learning Koopman and Input Matrix}\label{alg:1}
\begin{algorithmic}[1]
\Statex \textbf{Input: } \text{X,Y training data, batch size \(b_s,\) Epoch\(_{max}\)}
\Statex \textbf{Method}
\For{\text{epoch in range(Epoch\(_{max}\))}}
\For{\text{batch in range(number of batches)}}
\State \text{Sample \(b_s\) from training, control and label}

\Statex \hspace{\algorithmicindent} \hspace{\algorithmicindent} \text{data.}

\Statex \hspace{\algorithmicindent} \hspace{\algorithmicindent} \text{\(X = \{X^i_{0:k-1}\}^{b_s}_{i=1}, Y = \{Y^i_{1:k}\}^{b_s}_{i=1}\)}
\State \text{Encode the training and label data and stack}

\Statex \hspace{\algorithmicindent} \hspace{\algorithmicindent} \text{the original states onto the encoded states.}

\Statex \hspace{\algorithmicindent} \hspace{\algorithmicindent} \text{\(\Phi_x = [X; \Phi(X)]^T, \Phi_y = [Y; \Phi(Y)]^T\)}

\State \text{Compute the K matrix using EDMD.}

\Statex \hspace{\algorithmicindent} \hspace{\algorithmicindent} \text{\(K = \Phi_y \cdot (\Phi_x)^\dagger\)}
\State \text{Compute the next state with the linear}

\Statex \hspace{\algorithmicindent} \hspace{\algorithmicindent} \text{equation.}

\Statex \hspace{\algorithmicindent} \hspace{\algorithmicindent} \text{\(\hat{\Phi}_{x+1} = K\Phi_x\)}
\State \text{Apply the loss function \(\mathcal{L}\) using MSE}
\State \text{Update the model weights and parameters}
\Statex \hspace{\algorithmicindent} \hspace{\algorithmicindent} \text{with the Adam optimizer.}
\EndFor
\EndFor
\Statex \textbf{Output}: \text{K matrix, trained DNN}
\end{algorithmic}
\end{algorithm}

For any non-control affine system, the nonlinear dynamics are propagated by Equation \ref{eqn2}. After lifting the states to higher dimension with the observable function, $\boldsymbol{\Phi}$, we desire to find a linear system representation:

\begin{equation}
    \mathbf{\Phi}(\boldsymbol{x}_{k+1}) \approx \mathbf{K} \boldsymbol{\Phi}(\boldsymbol{x}_k),
\end{equation}

where the matrix $\mathbf{K}$ approximates the Koopman operator. This matrix is analogous to $\mathbf{A}$, the linear state transition matrix. The goal of the deep learning framework (Fig. \ref{fig:net}) is to learn \(N\) observables and calculate the $\mathbf{K}$ matrix through the appropriate construction of the loss functions. Because we need to approximate an infinite dimensional operator, we choose \(N\)  such that \(N >> n\) is satisfied, and define

\begin{equation}\label{eqn:observ}
    \mathbf{\Phi}(\boldsymbol{x}_k) := 
    \begin{bmatrix} 
    \boldsymbol{x}_k \\
    \phi_1(\boldsymbol{x}_k) \\ \phi_2(\boldsymbol{x}_k) \\ \vdots \\ \phi_N(\boldsymbol{x}_k) 
    \end{bmatrix},
\end{equation}

where \(\phi_i : \mathbb{R}^n \rightarrow \mathbb{R}, i = 1,...,N\) are the observable functions, learned by the DNN. An important observation is that we take the observables from the DNN and concatenate the original states \(\mathbf{x}_k\) on top, helping in two ways: (1) the original states can be easily extracted from the new set of observables (see below), and (2) valuable computational resources can be saved since there is no need for a decoder to recover the original states.

Currently, there is no generalized method to govern the size of \(N\) that would guarantee the optimal balance between simplicity and accuracy in the approximation of the Koopman operator; therefore, most often, \(N\) is chosen empirically through trial and error. 
There has been an increase in research that investigates ways to determine the size of \(N\) in both controllable and uncontrollable systems \cite{Zinage2023-vz, Lusch_2018}. In this work, DNNs are applied with EDMD to approximate the Koopman operator. 

To calculate the approximate Koopman operator $\mathbf{K}$, the time history of measurement data for \(M\) steps is arranged into snapshot matrices. The first snapshot matrix, \(X\), is the state history from time \(k=1\) to \(k=M-1\), whilst the second snapshot, \(X'\) is exactly the same as \(X\) but shifted forward one time-step: 

\begin{equation}
    X = \begin{bmatrix} \boldsymbol{x}_{1},  \boldsymbol{x}_{2},  \boldsymbol{x}_{3},  \dots, \boldsymbol{x}_{M-1} \end{bmatrix} 
\end{equation}
\begin{equation}
    X' = \begin{bmatrix} \boldsymbol{x}_{2},  \boldsymbol{x}_{3}, \boldsymbol{x}_{4},  \dots, \boldsymbol{x}_{M} \end{bmatrix} 
\end{equation}

Mapping the measured states, \(X, X'\) with observable functions leads to

\begin{equation}
    \boldsymbol{\Phi}(X) = \begin{bmatrix} 
    \boldsymbol{\Phi}(\boldsymbol{x}_1), 
    \boldsymbol{\Phi}(\boldsymbol{x}_2),  \dots, 
    \boldsymbol{\Phi}(\boldsymbol{x}_{M-1}) \end{bmatrix} 
\end{equation}
\begin{equation}
    \boldsymbol{\Phi}(X') = \begin{bmatrix} 
    \boldsymbol{\Phi}(\boldsymbol{x}_2), 
    \boldsymbol{\Phi}(\boldsymbol{x}_3),  \dots, 
    \boldsymbol{\Phi}(\boldsymbol{x}_M)  \end{bmatrix} 
\end{equation}    

Given the dataset, the matrix $\mathbf{K}$ can be found by using the least-squares method to minimize the following:

\begin{equation}
\min \sum\left\|\boldsymbol{\Phi}(\boldsymbol{x}_{k+1})-\mathbf{K} \boldsymbol{\Phi}\left(\boldsymbol{x}_k\right)\right\|^2
\end{equation}

Applying the snapshot matrices of real data yields

\begin{equation}
    \boldsymbol{\Phi}(X') \approx 
    \mathbf{K} \boldsymbol{\Phi}(X)
\end{equation}

therefore, on inversion, we end up with

\begin{equation}
        \mathbf{K} = \boldsymbol{\Phi}(X')\boldsymbol{\Phi}(X)^\dagger
\end{equation}

where the symbol $\dagger$ denotes the Moore-Penrose inverse of the matrix \cite{penrose_1955}. 

Because the Koopman operator calculated in this approach represents an approximation to the real, infinite dimensional operator, and as the observable functions do not span a Koopman invariant subspace, the predicted state is denoted with a circumflex (hat) symbol to emphasize that it is an approximation:

\begin{equation} \label{Kprop}
    \hat{\mathbf{\Phi}}(\boldsymbol{x}_{k+1}) = \mathbf{K} \hat{\mathbf{\Phi}}(\boldsymbol{x}_k)
\end{equation}

Now, we can extract the original states from the observables using a projection matrix $\mathbf{P}$ \cite{Junker_2022}, yielding

\begin{equation} \label{eqn:extract}
    \boldsymbol{x}_{k+1} = \mathbf{P}\hat{\mathbf{\Phi}}(\boldsymbol{x}_{k+1})\; \text{with}\; \mathbf{P} = \begin{bmatrix}
        \mathbf{I}_n, \mathbf{0}_{n \text{x} N}
    \end{bmatrix},
\end{equation}

where \(\mathbf{I}_n\) is the \(n \times n\) identity matrix and \(\mathbf{0}_{n \times N}\) is the \(n\times N\) zero matrix. As shown in \cite{Junker_2022} and \cite{Korda_2018}, the observable functions not spanning a Koopman invariant subspace accumulate error over time, which leads to predictions with high inaccuracy. However, this error can be mitigated if the prediction is corrected at each time step. This correction is applied by extracting the estimated state variable $\hat{x}_{k+1}$ at each time step with Equation \ref{eqn:extract}, and then reapplying the observable lifting functions to the extracted state variable with Equation \ref{eqn:observ}. 

\subsection{Orbital Dynamics}

One of the two main applications of the proposed model in this work is to linearize both the circular and elliptical 2BP orbit. Orbiting satellites in Low Earth Orbit (LEO) range in altitude from 200-300 km to around 2000 km \cite{pod_orbit}, and have an eccentricity of less than 0.25, whilst Geostationary orbiting satellites have an ideal eccentricity of 0 \cite{GEO_orbits}. With many of the operational satellites having an eccentricity less than 0.5 or very close to zero, this aspect motivates us to investigate the linearization of those orbits that are within this eccentricity range.  The equations of motion for a satellite in any orbit is a second-order differential equation that relates the movement of the satellite to the accelerations that perturb the orbit. These accelerations can depend on a number of parameters such as time \(t\), position \(\vec{\mathbf{r}}\), velocity \(\vec{\mathbf{v}}\) or physical forces \(\vec{\mathbf{f}}\). 

The equation of motion for a two-body orbit is given as follows:

\begin{equation} \label{orbdyn}
    \vec{\ddot{\mathbf{R}}}_2 - \vec{\ddot{\mathbf{R}}}_1 = -G(M+m)\frac{\vec{\mathbf{R}}_2 - \vec{\mathbf{R}}_1}{\left\|\vec{\mathbf{R}}_2 - \vec{\mathbf{R}}_1\right\|^3} + \vec{\mathbf{a}}
\end{equation}

Where, \(G\) is the Gravitational Constant, \(M\) is the mass of the central body, \(m\) is the mass of the orbiting body (satellite), \(\vec{\mathbf{R}}_2 - \vec{\mathbf{R}}_1\) is the distance from center of the satellite to the central body and \(\mathbf{a}\) is the additional perturbing acceleration. In our work, we set \(\mathbf{a} = \vec{\mathbf{0}}\) when dealing with purely circular or elliptical orbits to preserve the simplicity of the orbit, which should improve the DNN's ability to learn the observables, consequently leading to a more accurate representation of the Koopman operator. We also set the position of the central body, \(\vec{\mathbf{R}}_1\), to be at the origin, hence simplifying \ref{orbdyn} to:

\begin{equation}
    \vec{\ddot{\mathbf{r}}} = \mu\frac{\vec{\mathbf{r}}}{\left|\vec{\mathbf{r}}\right|^3}
\end{equation}

where \(\mu = -G(M+m)\) is defined as Earth's gravitational parameter. Note that because \(M >> m\), the mass of the satellite is often ignored in the calculation of \(\mu\); however, in our work, it is still taken into account and the abbreviation is held as such.

To prove the ability of the model to also handle some of these perturbing accelerations, we perform simulations which include orbital perturbations such as solar radiation pressure and a complex gravity model such as the J2 perturbation. The models for both perturbations can be found in \cite{schaub} and for these simulations, the nonlinear dynamics can be represented as follows:

\begin{equation}
    \vec{\ddot{\mathbf{r}}} = \mu\frac{\vec{\mathbf{r}}}{\left|\vec{\mathbf{r}}\right|^3} + \vec{\mathbf{a}}_{J2} + \vec{\mathbf{a}}_{SRP}
\end{equation}

We choose to represent the orbits as in-plane, thus in two dimensions \cite{MD99,vallado1997fundamentals}, resulting in \(z=0\) for all time. This results in four states being required to represent the nonlinear system in state space:

\begin{equation}
    \mathbf{x} = \begin{bmatrix} x_1 & x_2 & x_3 & x_4
    \end{bmatrix}^{\mkern-1.5mu\mathsf{T}} = 
    \begin{bmatrix} x & y & \dot{x} & \dot{y}
    \end{bmatrix}^{\mkern-1.5mu\mathsf{T}}
    \label{orbit eom}
\end{equation}

And the state space representation of the nonlinear model would correspond to:

\begin{equation}
\begin{aligned}
    \dot{\mathbf{x}} &= \begin{bmatrix} \dot{x_1} \\ \dot{x_2} \\ \dot{x_3} \\ \dot{x_4}
    \end{bmatrix} = 
    \begin{bmatrix} \dot{x} \\ \dot{y} \\ \ddot{x} \\ \ddot{y}
    \end{bmatrix} \\ &= 
    \begin{bmatrix} 0 & 0 & 1 & 0 \\ 0 & 0 & 0 & 1 \\ \frac{\mu}{\left|\vec{\mathbf{r}}\right|^3} & 0 & 0 & 0 \\ 0 & \frac{\mu}{\left|\vec{\mathbf{r}}\right|^3} & 0 & 0
    \end{bmatrix}
    \begin{bmatrix} x_1 \\ x_2 \\ x_3 \\ x_4
    \end{bmatrix}
    \label{orbit ss}
\end{aligned}
\end{equation}

To ease the generation of data for this system, each orbit obtains the same initial pose, chosen to be at the periapsis of the orbit, \(x=r_p,\: y=0\). Hence, the initial velocity is purely tangential, and defined as:

\begin{equation}
\begin{aligned}
    & \dot{x} = 0 \\
    & \dot{y} = \sqrt{\mu\left(\frac{2}{r_p}-\frac{1}{a}\right)}
\end{aligned}
\label{2BPvel}
\end{equation}

Where, $r_p$ is the radius of periapsis for the orbit and $a$ is the semi-major axis. For purely circular orbits this simplifies to: $ \dot{y} = \sqrt{\frac{\mu}{r_p}}$.

Hence, the initial condition for any orbit is as follows:

\begin{equation}
    \mathbf{x_0} = \begin{bmatrix}
        r_p & 0 & 0 & \sqrt{\mu\left(\frac{2}{r_p}-\frac{1}{a}\right)}
    \end{bmatrix}^{\mkern-1.5mu\mathsf{T}}
\label{2BPic}
\end{equation}

and $r_p$ is a randomly selected parameter for each initial condition in the training dataset. 

\subsection{Circular Restricted Three-Body Dynamics} \label{CR3BPdyn}

As previously mentioned, the periodic orbits of the CR3BP are often derived around one of the five Lagrange points. One of these Lagrange points (L1) will be the case utilized in this work to generate the data required for training, in particular, orbits that oscillate and are influenced by the L1 Lagrange point.
In the application of the CR3BP, the mass of the satellite is considered to be negligible relative to the primary \(P_1\) and secondary \(P_2\) masses. The propagation of the CR3BP dynamics also requires that the mass, length and time units be non-dimensionalized by the following factors:

\begin{equation}
    \begin{aligned}
        M^* &= M_E + M_M \\
        L^* &= a \\
        T^* &= \sqrt{\frac{L^{*3}}{GM*}}
    \end{aligned}
\end{equation}

where \(M_E\) and \(M_M\) are the masses of the Earth and Moon respectively with \(M_M < M_E\), \(G\) is the universal gravitational constant, and \(a\) is semi-major axis of the system (distance between both primary bodies). We can now define the mass fraction \(\mu\) as the ratio between the secondary mass to the total mass:

\begin{equation}
    \mu = \frac{M_M}{M^*}
\end{equation}

For the propagation of the dynamics in the state space, a 6-dimensional state vector \(\mathbf{x}\) is implemented:

\begin{equation*}
    \mathbf{x} = \begin{bmatrix}
        x & y & z & \dot{x} & \dot{y} & \dot{z}
    \end{bmatrix}^{\mkern-1.5mu\mathsf{T}}
\end{equation*}

where \(x, y, z\) are the positions in the rotating frame and the dot indicates the derivative with respect to the non-dimensional time (i.e., normalized by $T^*$). This reference frame has its origin at the barycenter of the primary and secondary masses, and has its x axis aligned in the direction to the secondary mass.

The equations of motion describing the motion of the third mass \(P_3\) can now be derived as:

\begin{equation}
    \begin{aligned}
        \ddot{x} &= x + 2\dot{y} - \frac{(1-\mu)(x+\mu)}{d^3} - \frac{\mu(x-1+\mu)}{r^3} \\
        \ddot{y} &= y - 2\dot{x} - \frac{y(1-\mu)}{d^3} - \frac{\mu y}{r^3} \\
        \ddot{x} &= - \frac{(1-\mu)z}{d^3} - \frac{\mu z}{r^3}
    \end{aligned}
\end{equation}

Where the following substitutions are made:

\begin{equation}
    \begin{aligned}
        d &= \sqrt{(x+\mu)^2+y^2+z^2} \\
        r &= \sqrt{(x-1+\mu)^2+y^2+z^2}
    \end{aligned}
\end{equation}

Due to the fact that we are interested in learning a family of periodic orbits that oscillate around the L1 Lagrange point, the initial conditions for the orbit need to be selectively chosen. The orbit is in the orbital plane, so we can inherently assume the $z$-axis position and velocity values to be zero. The initial $y$ position is given as \(\frac{1}{a}\), whilst the initial $x$ position is found as the L1 Lagrange point by solving the following equation for the value of \(x\) bounded by the position of the Earth and the Moon:

\begin{equation} \label{CR3BPIC}
    \begin{aligned}
        0 = &-\frac{1-\mu}{(x+\mu)\cdot\left|x+\mu\right|} -\frac{\mu}{(x-1+\mu)\cdot\left|x-1+\mu\right|} + x
    \end{aligned}
\end{equation}

The $x$- and $y$-components of the velocity may now be calculated by making use of the initial positions, the Lagrange point around which oscillations occur, and the mass fraction of the system, through a number of complex equations that are described in \cite{CR3BPvel}. In order to generate a number of accurate simulations, the \textit{cr3bp} Python package is utilized to find these initial conditions following the same procedure.

In order to produce trajectories that can be used in training with enough variance to adequately train the DNN, the initial position in the $x$ direction was multiplied by a random number in the range \([1, 1.05]\). Such a small range is required since the nature of the orbit and its periodic behavior are highly sensitive to the initial conditions, a manifestation of chaotic behavior. Multipliers of any value outside of this narrow range would result in orbits that were not periodic around the L1 point.

\section{NEURAL NETWORK ARCHITECTURE}
\subsection{Deep Neural Network Structure and Loss Function Definition}

The development of DNN's has been aided through the improvement in support packages such as PyTorch and Keras TensorFlow. Whilst these packages are often equipped with predetermined loss functions for regression and classification problems, the general use of a DNN requires a custom loss function to optimize the performance of the network to the particular application. Our approximation of the Koopman operator is achieved through a DNN designed with a custom loss function that is a summation of multiple loss functions that are integral for accurate dynamic propagation prediction. The DNN structure used in this work is a fully connected, feedforward network, as it is the most fundamental architecture, and is often used in regression and prediction problems. Each hidden layer has a scaled exponential linear unit (SELU) activation function, whilst the final output layer has no activation function. 

The DNN learns the lifting observable functions which are in turn, used to determine the approximation to the Koopman operator through EDMD. In order to learn these observable functions, we design the following loss functions that are calculated using a recursive mean square error (MSE) formula:
\begin{equation} 
\begin{aligned}
	\mathcal{L}_{\text {Recon}} &=\frac{1}{N_d} \sum_{k=1}^{N_d}\left\|{\hat{\boldsymbol{x}}_{k+1}}-{\boldsymbol{x}_{k+1}} \right\|_2^2 \\
	\mathcal{L}_{\text {Pred}} &=\frac{1}{N_{pred}} \sum_{k=1}^{N_{pred}}\left\|{\hat{\boldsymbol{x}}_{k+\alpha}}-{\boldsymbol{x}_{k+\alpha}} \right\|_2^2 \\
\end{aligned}
\label{lossfcn}
\end{equation}

where \(\mathcal{L}_{\text {Recon}}\) represents the one-time-step reconstruction loss function that has the objective of ensuring proper reconstruction of the original states at each time step. In contrast, \(\mathcal{L}_{\text {Pred}}\) is our custom prediction loss function that is used to improve the consistency of a prediction \(\alpha\) time steps into the future. \(\alpha\) is a user defined parameter, set during the data generation stage allowing the DNN to ensure that it is able to correctly evolve multiple time steps into the future. The benefit of this loss function is that it is able to decrease the drift of the predicted trajectory from the original nonlinear dynamical system by allowing the network to learn multi-step predictions. The state \(\boldsymbol{\hat{x}_{k+\alpha}}\) is calculated using a \emph{for} loop of length \(\alpha\), where each next observable is calculated using the linear dynamics \(\hat{\mathbf{\Phi}}(x_{k+1}) = \mathbf{K} \hat{\mathbf{\Phi}}(x_k)\) and the original-dimension state is extracted using Equation \ref{eqn:extract}. 

In order to further improve the accuracy of the model learned, training techniques such as \(L_1\) and \(L_2\) regularization are implemented into the loss function as a means to penalize unwanted behaviors by the DNN \cite{L1vsL2}. For instance, \(L_1\) regularization aims to penalize the model for having large weights, achieving this objective by including the sum of the absolute value of the weights in the loss function, thus calculated as:

\begin{equation}
    \mathcal{L}_{L_1}=\lambda_1\sum_{k=1}^{N_w}|w_k|
\end{equation}

where \(\lambda_1\) is the weighting factor applied to the \(L_1\) regularization, \(N_w\) is the number of total weights in the entire network, and \(w_k\) is the $k$-th weight parameter. The result of this regularization is that features with lesser importance have their weights reduce to near-zero or zero, removing them from the approximation completely. Naturally, because of the way \(L_1\) regularization reduces the unimportant weights to zero, the resulting network is sparse. Hence, this regularization constitutes a good pairing with EDMD, since EDMD seeks to extract only the most important information from the dynamics. 

\(L_2\) regularization on the other hand, encourages the sum of the squares of the weights to be small, thus reducing the chance of overfitting, whilst improving the networks ability to learn complex features. This is particularly useful when handling nonlinear dynamics of increased dimension and increased terms as in the CR3BP. The \(L_2\) regularization loss function can be calculated as:

\begin{equation}
    \mathcal{L}_{L_2}=\lambda_2\sum_{k=1}^{N_w}(w_k)^2
\end{equation}

Lastly, because we have knowledge of the system that we desire to represent, we can make use of some physics-informed loss functions to further help our network model the physical properties of the system as accurately as possible. Section \ref{2BPmetrics} explains the accuracy metrics that are used to ensure physical consistency between our model and the exact nonlinear dynamics however for the 2BP model, we utilize the third metric, $r \cdot v$, as an additional loss function. We again use the MSE formula to generate the loss, and because we desire this metric to be constantly zero, we use that as our truth to remove numerical errors:

\begin{equation}
    \mathcal{L}_{r \cdot v} = \frac{1}{N_d}\sum_{k=1}^{N_d}\left\|{\hat{(r \cdot v)}-0}\right\|_2^2
    \label{rdotvlossfcn}
\end{equation}

We can now construct the total loss function as:

\begin{equation}
    \mathcal{L}_{\text {total}}=\gamma\mathcal{L}_{\text {Pred}}+\beta\mathcal{L}_{\text {Recon}} + \lambda_1\mathcal{L}_{L_1} + \lambda_2\mathcal{L}_{L_2} + \lambda_{r \cdot v}\mathcal{L}_{r \cdot v}
\end{equation}

The total loss function value, \(\mathcal{L}_{\text {total}}\) is then used in the Adam optimizer, to recursively improve the weights and biases of the DNN, so that the observable functions can be appropriately learned. With improved observable functions, the approximation to the Koopman operator becomes more accurate, hence the performance of the linear system becomes more true. The weights \(\gamma, \beta, \lambda_1, \lambda_2, \lambda_{r \cdot v}\) corresponding to each loss function, can be adjusted according to the performance of the network and knowledge of the dynamics of the system. 

\section{MODEL ACCURACY METRICS}

In order to prove the accuracy and efficacy of an algorithm or model, it is important to benchmark the results produced against some metrics relevant to the problem. In this section, we present a number of metrics that apply to either the 2BP or CR3BP that allow us to determine the accuracy of our model. By considering physical constants of each orbit type, we can determine if our model truly captures the underlying dynamics of the system. 

\subsection{Global vs Local Error}\label{errors}

In dynamic prediction there is a distinction between the global error in prediction and the local error. While the local error, $e_n$, only considers the error between the predicted model and the exact model over one time-step, the global error, $E_n$ considers the error over all time-steps and is a more accurate measure of model performance \cite{10091950}. An important distinction is the fact that the global error is not equivalent to the summation of all local errors, this is shown in \cite{10091950}. The definitions for both errors are given below, and we highlight our model performance by displaying both errors for all simulations in the 2BP problem. 

\begin{equation}
    e_n \equiv \mathbf{\Phi}(\boldsymbol{x}_n)- \boldsymbol{K}\mathbf{\Phi}(\boldsymbol{x}_{n-1})
\end{equation}

\begin{equation}
    E_n \equiv \mathbf{\Phi}(\boldsymbol{x}_n)- \boldsymbol{K^n}\mathbf{\Phi}(\boldsymbol{x}_{0}) \equiv \mathbf{\Phi}(\boldsymbol{x}_n)- \boldsymbol{K}\hat{\mathbf{\Phi}}(\boldsymbol{x}_{n-1})
\end{equation}

where the recursive nature of the estimated state $\hat{\mathbf{\Phi}}$ is defined in Equation \ref{Kprop}.

\subsection{Two-Body Circular Metrics} \label{2BPmetrics}
The \emph{instantaneous} coordinates of a circular orbit at some time $t$ are $x(t)$ and $y(t)$. The operator `$\cdot$' denotes the time derivative; for example, $\dot{x} \equiv dx/dt$ is the $x$-component of the velocity. We introduce the notation:
\begin{equation}
    {\bf r}(t) = \left(x(t),y(t)\right),
\end{equation}
\begin{equation}
    {\bf v}(t) = \left(\dot{x}(t),\dot{y}(t)\right),
\end{equation}
where ${\bf r}$, ${\bf v}$, and ${\bf a}$ signify the position, velocity, and acceleration vectors, respectively for motion in a plane (i.e., two-dimensional motion). 

During the course of an entire orbit, let us suppose that $N$ measurements are performed at the times $t = t_1,\, t_2,\, \dots t_N$. We shall define an orbit average (i.e., the average over a single orbit) $\langle{\chi}\rangle$ of a given time-dependent quantity $\chi(t)$ -- which can represent either a scalar, vector, or tensor -- as follows:
\begin{equation}\label{OrbAvg}
    \langle{\chi}\rangle = \frac{1}{N} \sum_{i = 1}^N \chi\left(t_i\right).
\end{equation}
Note that $\langle{\chi}\rangle$ is not a function of time, since it is a temporal average computed over all times (in the specific context of a single orbit).

Once the orbit average is determined, we can estimate the relative variation $\xi_\chi(t)$ of the aforementioned $\chi(t)$ -- which constitutes a useful measure of the relative error for invariants of circular orbits -- through the expression,
\begin{equation}\label{RelErrDef}
    \xi_\chi(t) = \frac{\chi(t)}{\langle{\chi}\rangle} - 1.
\end{equation}
It is important to recognize a couple of properties associated with $\xi_\chi(t)$: (a) it can be either positive or negative, although $|\xi_\chi(t)|$ will always be positive by definition; (b) for an invariant of the system, we would ideally have $\chi(t) = \mathrm{const} = \langle{\chi}\rangle$, implying that $\xi_\chi(t) \equiv 0$. However, when numerical algorithms are utilized, $\chi(t)$ may not be perfectly invariant, owing to which $\xi_\chi(t)$ could thus be non-zero, albeit small in magnitude for robust algorithms.

In circular motion, the following quantities are all conserved, i.e., they are invariants of circular motion.

\begin{enumerate}
    \item $r \equiv |{\bf r}| = \sqrt{x^2 + y^2}$ - The orbital radius, $r$, of a circular orbit should remain constant throughout the entire orbit. This constitutes the definition of a circular orbit.
    \item $v \equiv |{\bf v}| = \sqrt{\dot{x}^2 + \dot{y}^2}$ - The magnitude of velocity, $v$ of a circular orbit should remain constant throughout the entire orbit. 
    \item ${\bf r} \cdot {\bf v} = x \dot{x} + y \dot{y}$ - The velocity vector in a circular orbit should remain perpendicular to the radial direction, hence their dot product should be zero.
    \item $L_z \equiv \left({\bf r} \times {\bf v}\right)_z = x \dot{y} - y \dot{x}$ - The angular momentum, $L_z$, of a circular orbit, defined as the cross product of the orbital radius with the tangential velocity should be a constant value.
\end{enumerate}
In addition, the Kepler problem is endowed with invariants such as the \\Laplace–Runge–Lenz vector that will not be addressed in this work \cite{BNA23}.

For each of the above invariants, except for \#3, we can construct their corresponding relative error. For instance, in the case of $r$, the relative error $\xi_r(t)$ is given by
\begin{equation}
      \xi_r(t) = \frac{r(t)}{\langle{r}\rangle} - 1
\end{equation}
after invoking (\ref{RelErrDef}); where $r(t) = \sqrt{x^2(t) + y^2(t)}$, and $\langle{r}\rangle$ is specified by (\ref{OrbAvg}). As mentioned in the paragraph below (\ref{RelErrDef}), it is expected that $\xi_\chi(t) \equiv 0$ for invariants of the system. Therefore, given that $r$ is supposed to be an invariant of circular motion, the numerical algorithm(s) employed should yield values of $\xi_r(t)$ close to zero; in fact, the amount of deviation from zero is a good rubric for evaluating the accuracy of the deployed algorithm(s).

From the determination of the third metric (namely \#3), which is ${\bf r} \cdot {\bf v}$, in the circular 2BP this value should be equivalent to zero as the velocity vector should always be tangential to the positional vector, however in numerical simulation it is found that the value of this metric follows a small sinusoidal shape for one orbital period of calculation. Hence, when the average of the entire dataset is taken, this value tends toward zero, consequently -- when applied to the relative error formula given by (\ref{RelErrDef}) -- yields a substantially greater magnitude of error than what is actually observed in the simulation. Therefore, in the discussion of the ${\bf r} \cdot {\bf v}$ metric, it should be noted that the metric is not a relative error but a plot of the actual value. We can then determine the efficacy of this metric by determining how far the value is from zero at every calculated point. 

\subsection{CR3BP Jacobi Constant} \label{jacobi}

In order to determine the accuracy of the algorithm on predicting a true representation of the CR3BP, the aforementioned metrics would not suffice, because they are exclusively applicable to the circular 2BP. We can however, utilize the conserved quantity known as the Jacobi Constant, which is a relative negative measure of energy \cite{MD99,Jacobi1} to ensure whether or not our method is satisfactory. Given that the Jacobi constant is conserved, and it is the only constant of motion in the CR3BP (since it has no ignorable coordinates \cite{Lagrangian}),  hence it is a metric that we can employ on our predicted Koopman dynamics. The rigorous proof of the Jacobi constant can be derived in numerous ways \cite{MD99, battin1999introduction, vallado1997fundamentals}, however the derivation of the equation used in this work mirrors \cite{Lingam}, with the Jacobi constant defined as:

\begin{equation}
    C_j = \Omega^2r^2 + \frac{2\mu_1}{r_1} + \frac{2\mu_2}{r_2}-v^2
\end{equation}

in which \(\Omega\) is the uniform angular rotation of the barycenter of the CR3BP, \(r^2 = x^2 + y^2\) with \(x\) and \(y\) being the positions of the satellite in the rotating frame, \(\mu_1\) and \(\mu_2\) being the mass fractions of the primary and secondary masses respectively, and where \(r_1\) and \(r_2\) are the positions of the satellite relative to the primary and secondary masses respectively.

\section{SIMULATION AND RESULTS}

In this section, we present the simulation, results and discussion of the implementation of our proposed method. We perform a simulations of the pure circular and elliptical nonlinear dynamics of the 2BP, the SRP and J2 perturbed circular 2BP and the CR3BP around the L1 Lagrange point. We also perform numerical analysis of the aforementioned physical accuracy metrics for both the 2BP and the CR3BP. The full PyTorch code for data generation, training and the examples provided can be found on GitHub\footnote[1]{\url{https://github.com/tiwari-research-group/Orbital-Koopman}}. 

\subsection{Data Generation}

In order to adequately train the DNN, a large dataset of varying initial conditions (IC's) associated with orbital propagation is collected prior. The method in which we derive a range of different IC's for the training data is slightly different for both the variations of the 2BP and CR3BP, however the data generation method after the IC has been selected is the same for both dynamics -- refer to Algorithm \ref{alg:datagen}. 

\begin{algorithm}
\caption{Data Generation Algorithm}\label{alg:datagen}
\begin{algorithmic}[1]
\Statex \textbf{Input: } \text{Number of Initial Conditions \(n_{IC}\),}
\Statex \text{ time step \(dt\), final time \(t_f\), data points \(dp\).}
\Statex \textbf{Method}
\For{\text{num in range(\(n_{IC}\))}}

\State \text{Generate random I.C based on \ref{2BPic} or \ref{CR3BPIC}}

\Statex \hspace{\algorithmicindent} \text{\(x_{init} = x_{0,2BP}\) or \(x_{0,CR3BP}\)} 

\State \text{Declare the time vector for each I.C}

\Statex \hspace{\algorithmicindent} \text{\(time =[0 \sim t_f]\)}

\For {\text{t in time}}
\State \text{Save the current state in the X data array}

\State \text{Integrate the current state using the dynamics}
\Statex \hspace{\algorithmicindent} \hspace{\algorithmicindent}\text{and RK4 functions.}

\State \text{Save the new state in the Y data array.}
\EndFor
\State \text{Stack the solutions, X, Y from the I.C. into 3D}
\Statex \hspace{\algorithmicindent} \text{array that holds every trajectory for each I.C.}

\Statex \hspace{\algorithmicindent} \text{\(data_i\)}
\EndFor

\Statex \textbf{Output}: \text{Data matrices: \(data_x\), \(data_y\),}
\Statex \hspace{\algorithmicindent} \hspace{\algorithmicindent} \text{Input parameters: \(dt, t_f, dp, m\).}
\end{algorithmic}
\end{algorithm}

\subsubsection{Two-Body Data Generation}
In order to create either the 2BP circular or elliptical orbit, the correct IC's must be selected. The semi-major axis \(a\) and eccentricity \(e\) of an orbit are the only orbital elements needed to parameterize the shape of the orbit. Hence, only these will be calculated and used to determine the IC for the particular orbit. In particular the orbital element which will be randomly varied in the generation of the data is the semi-major axis \(a\). This is done implicitly by randomly choosing an altitude between 200 km and 5000 km and using the following relation:

\begin{equation}
    a = R_E + h
\end{equation}

where \(R_E = 6378.14 km\) is the radius of the Earth (assumed to be spherical) and \(h\) is the randomly chosen altitude. For the pure circular and perturbed circular orbits, we set $e=0$ whilst for the elliptical orbits the eccentricity is sampled from the range $[0.1-0.5]$. From this information, we can use equations \eqref{2BPvel} and \eqref{2BPic} to determine the initial positions and velocities for the current orbit. Each orbit is propagated for the length of one period (\(T = 2\pi\sqrt{\frac{a^3}{\mu}}\)) before being truncated by \(\alpha\) data points to accommodate the prediction loss function outlined in \eqref{lossfcn}. For the 2BP, \(\alpha = 25\). The number of IC's in each training set was 200 for the 2BP.

The generation of SRP in the perturbed orbits requires a Sun position vector, from which the direction of perturbation can be calculated. As the training data contains only one period of an orbit, it is short enough to assume that the position vector of the Sun is constant. We take the position of the Sun to be in the negative X-direction (i.e. $[-1,0]$).

\subsubsection{Circular Restricted Three-Body Data Generation}

As mentioned in Section \ref{CR3BPdyn}, the family of orbits that we adopt for our training data are periodic around the L1 Lagrange point, thus the range of IC for which we generate the training data is fairly small because of the high sensitivity the dynamics have to the IC. Section \ref{CR3BPdyn} and \cite{CR3BPvel} outline the method used for determining IC's that have a degree of randomness, necessary for training the DNN. Each orbit is not symmetric, nor exactly periodic, hence there is no analytical method to determine the length of a period, therefore an empirical length of 90 hours was chosen for each data set. This length of time, for the purposes of simulation, was non-dimensionalized with the \(T^*\) value derived for the dynamics. Like the 2BP, the data points collected was \(dp=1000\) and the truncation by \(\alpha\) was applied to each IC dataset. The value of \(\alpha\) was the same as the 2BP, however the total number of IC in each training set was 500.

\subsection{Results and Discussion}
\subsubsection{Two-Body Problem Linearization}

In this section we discuss the results of the proposed architecture on the application to the 2BP as well as the structure of the NN used to learn the approximate Koopman operator. For practicality purposes, the two-body system in which the Earth is the primary body is the main system considered for the following demonstration, although it is worth noting that the learned Koopman operator works well for other celestial circular two-body orbits, albeit with slightly larger error.

The hyperparameters of the DNN used for all experiments with the 2BP are outlined in Table \ref{table:1}. Total training time for the DNN used in the 2BP was 2 hours on an NVIDIA GeForce RTX 3090 GPU. The hyperparameters of both networks are tuned emperically, whilst it is noted that large networks (both in depth and width) do not necessarily result in improved results. 

\clearpage 

\begin{table}[h]
\begin{center}
\begin{tabular}{ |p{5cm}||p{6cm}| }
 \hline
 \multicolumn{2}{|c|}{\textbf{2BP Neural Network Hyperparameters}} \\
 \hline
 \textbf{Hyperparameter} & \textbf{Value} \\
 \hline
 Lifted Space Size   & 6\\
 Hidden Layers &   3\\
 Neurons per hidden layer & 25\\
 Batch Size & 128\\
 Learning Rate    &0.0001\\
 Optimizer &   Adam\\
 Activation Function & SELU\\
 Weight Decay & 0.00001\\
 Epochs & 80000\\
 \(\gamma\) & 0.8\\
 \(\beta\) & 1\\
 \(\lambda_{L_1}\) & 0.04\\
 \(\lambda_{L_2}\) & 0.01\\
 \(\lambda_{r \cdot v}\) & 0.001\\
 \hline
 K matrix dimension & 10 x 10\\
 \hline
\end{tabular}
\end{center}
\caption{Hyperparameters for 2BP Neural Network}
\label{table:1}
\end{table}

\begin{figure}[H]
    \centering
    \includegraphics[width = 0.85\columnwidth]{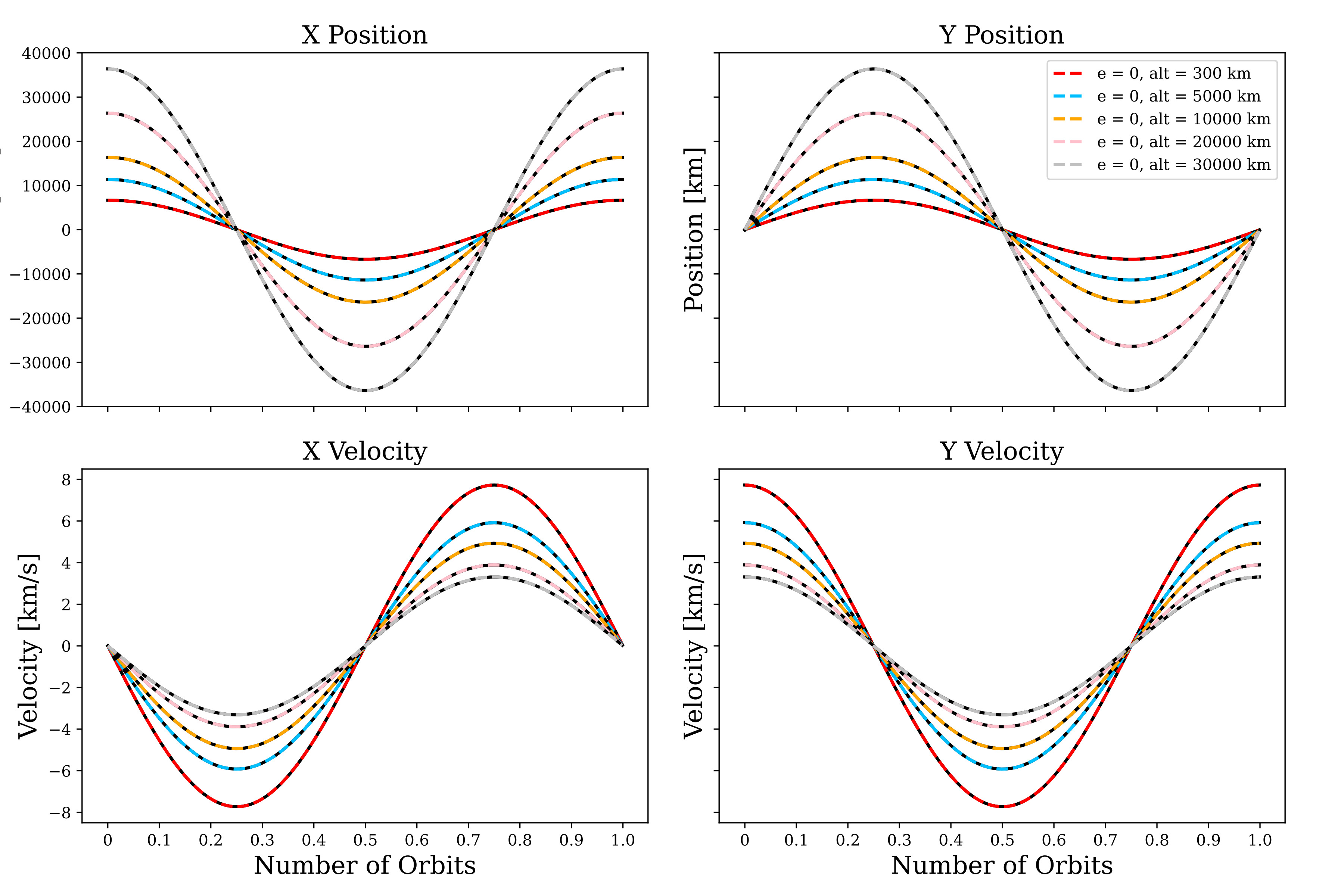}
    \caption{Four position and velocity states of the Earth orbits over one orbital period in comparison to the full nonlinear dynamics.}
    \label{fig:states_earth}
\end{figure}

\begin{figure}[H]
    \centering
    \includegraphics[width = 0.85\columnwidth]{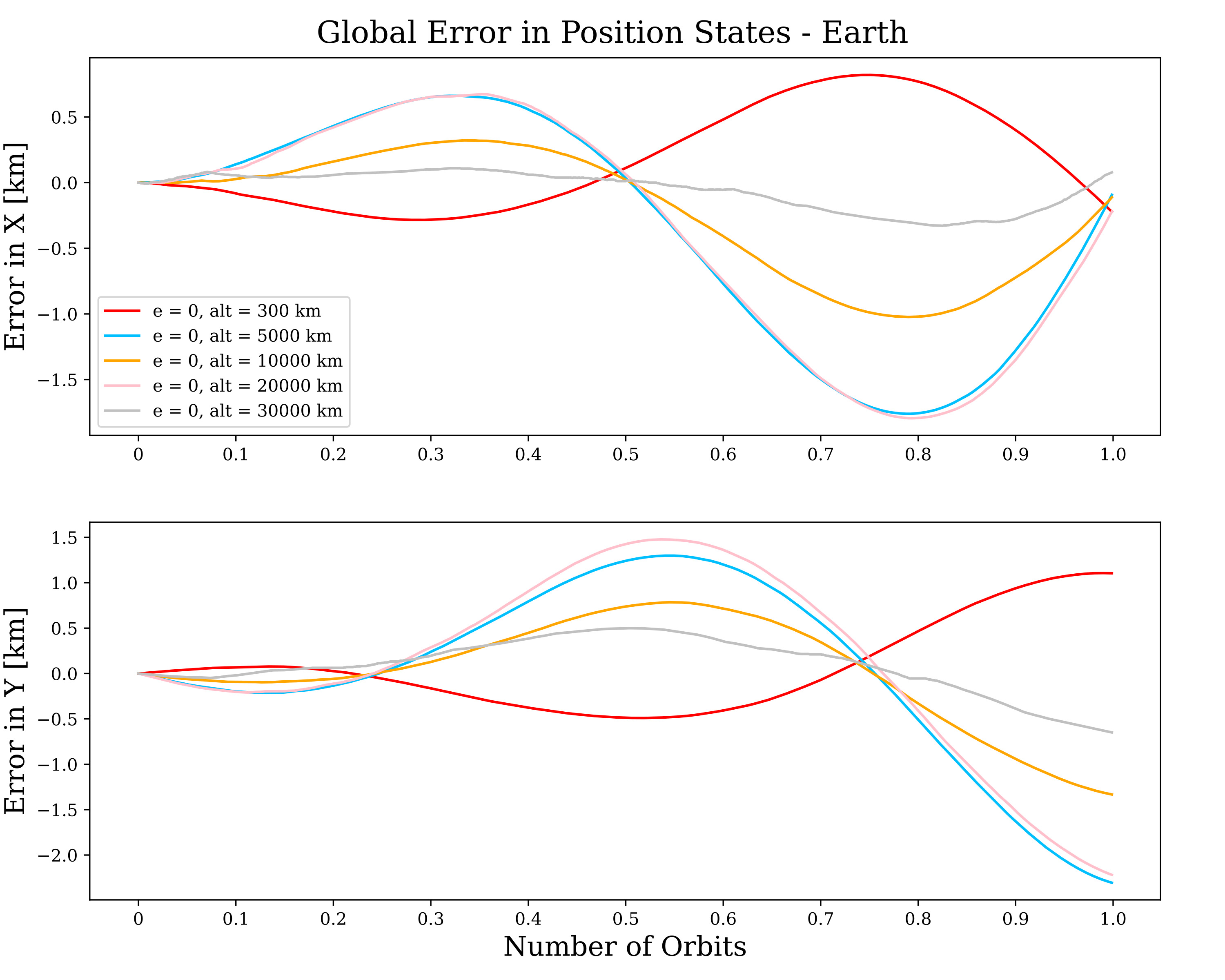}
    \caption{Global error in position states over one orbital period around Earth.}
    \label{fig:global_error_earth}
\end{figure}

\begin{figure}[H]
    \centering
    \includegraphics[width = 0.85\columnwidth]{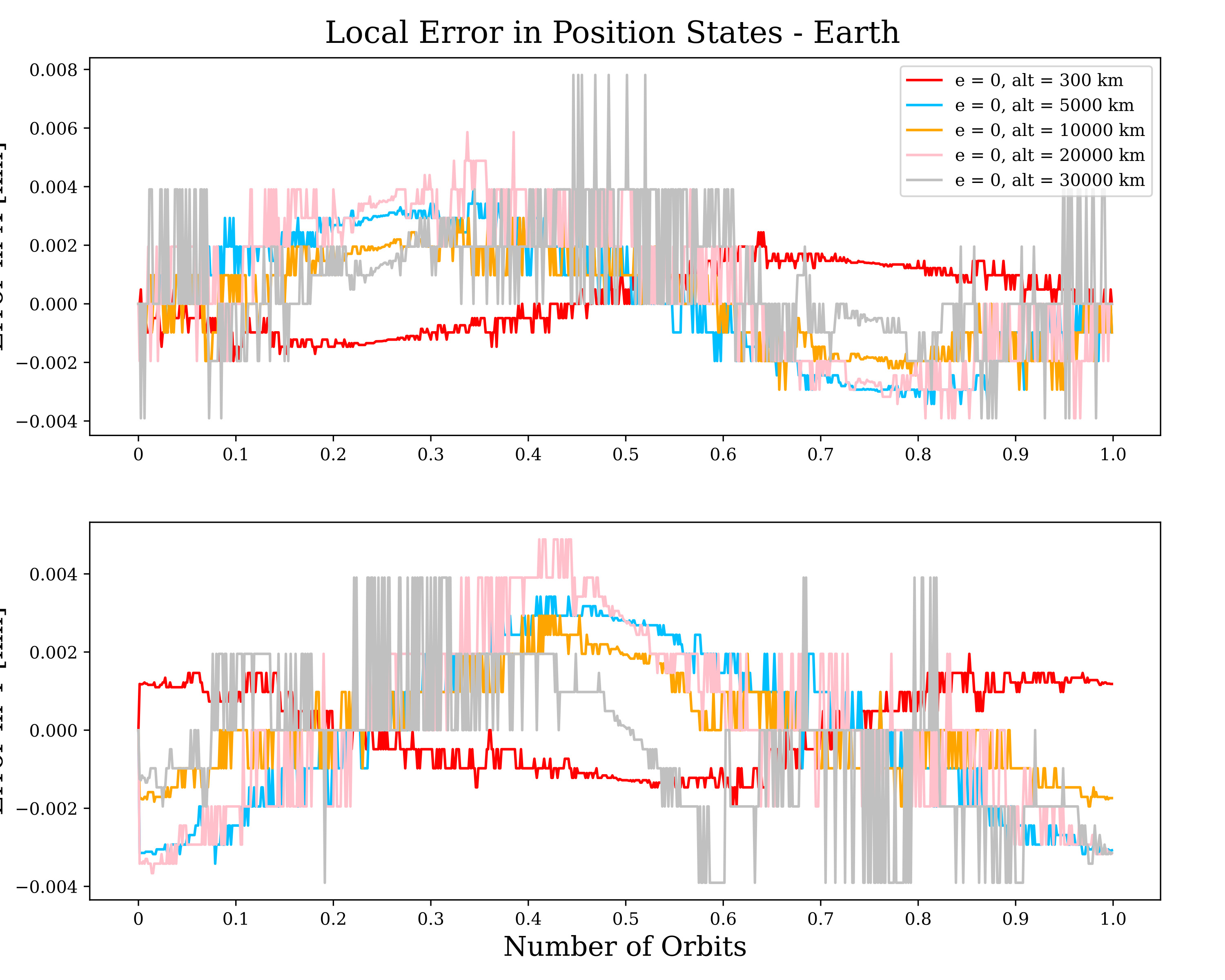}
    \caption{Local error in position states over one orbital period around Earth.}
    \label{fig:local_error_earth}
\end{figure}

Figure \ref{fig:states_earth} shows the prediction by the approximated Koopman operator on a range of purely circular orbits of varying semi-major axes. It can be noted that the prediction of the orbits are very accurate for regions both inside and outside the training region, aligning with the nonlinear dynamics very closely. Figures \ref{fig:global_error_earth} and \ref{fig:local_error_earth} display the global and local errors in prediction as outlined in Section \ref{errors}. The largest global positional errors in the single orbit prediction are on the order of 2 km, equivalent to $0.018\%$ of the radius of the orbit, whilst the local error has an maximum positional 8 m, around $2.16e^{-6}\%$ of the orbital radius. It is evident from Figure \ref{fig:states_earth} that for each varying altitude trajectory  not only is position, but the velocity of the satellite is also modeled quite well.

\begin{figure}[H]
    \centering
    \includegraphics[width = 0.85\columnwidth]{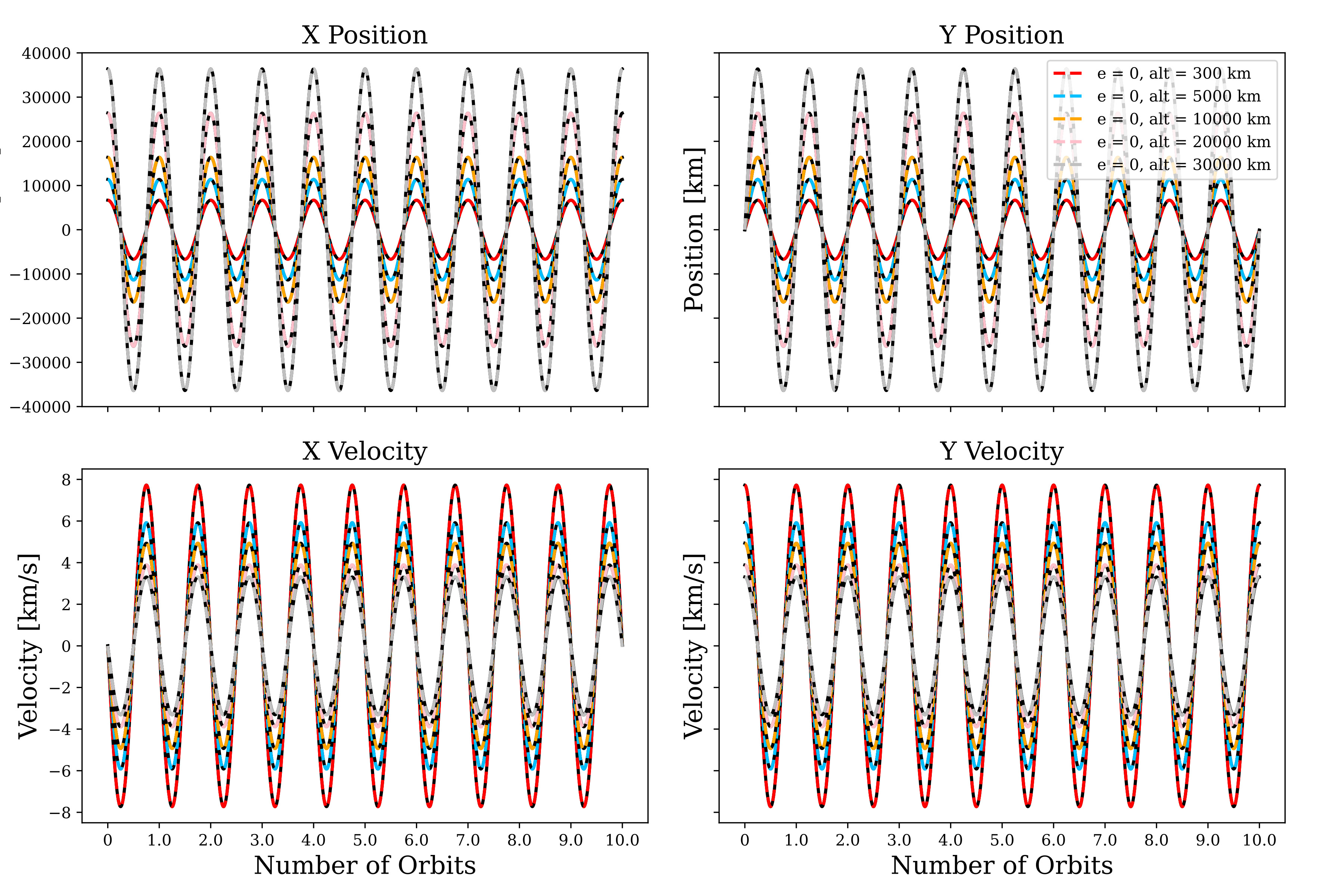}
    \caption{Four position and velocity states of the Earth orbits over ten orbital periods in comparison to the full nonlinear dynamics.}
    \label{fig:states_10}
\end{figure}

\begin{figure}[H]
    \centering
    \includegraphics[width = 0.85\columnwidth]{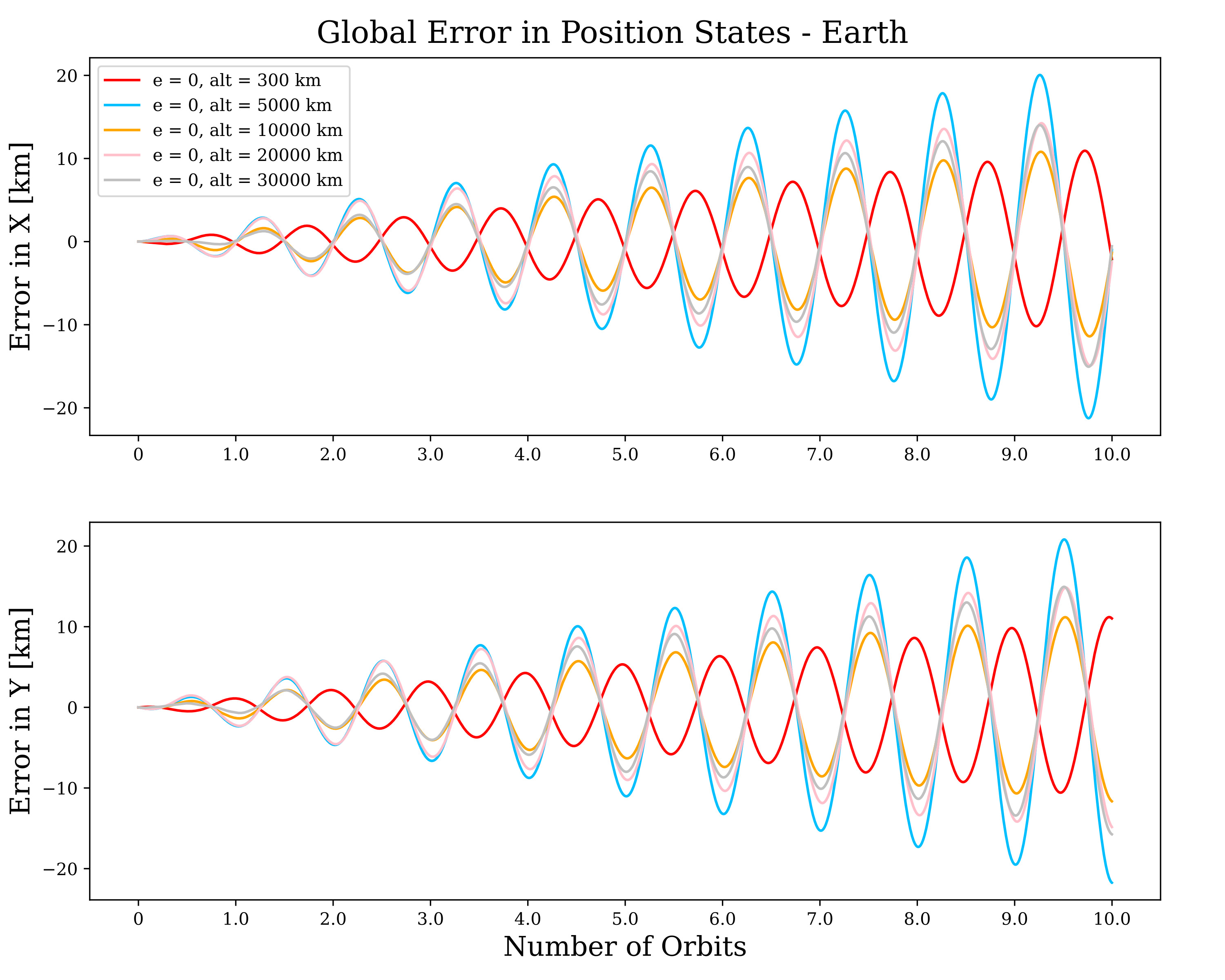}
    \caption{Global error in position states over ten orbital periods around Earth.}
    \label{fig:global_error_10}
\end{figure}

\begin{figure}[H]
    \centering
    \includegraphics[width = 0.85\columnwidth]{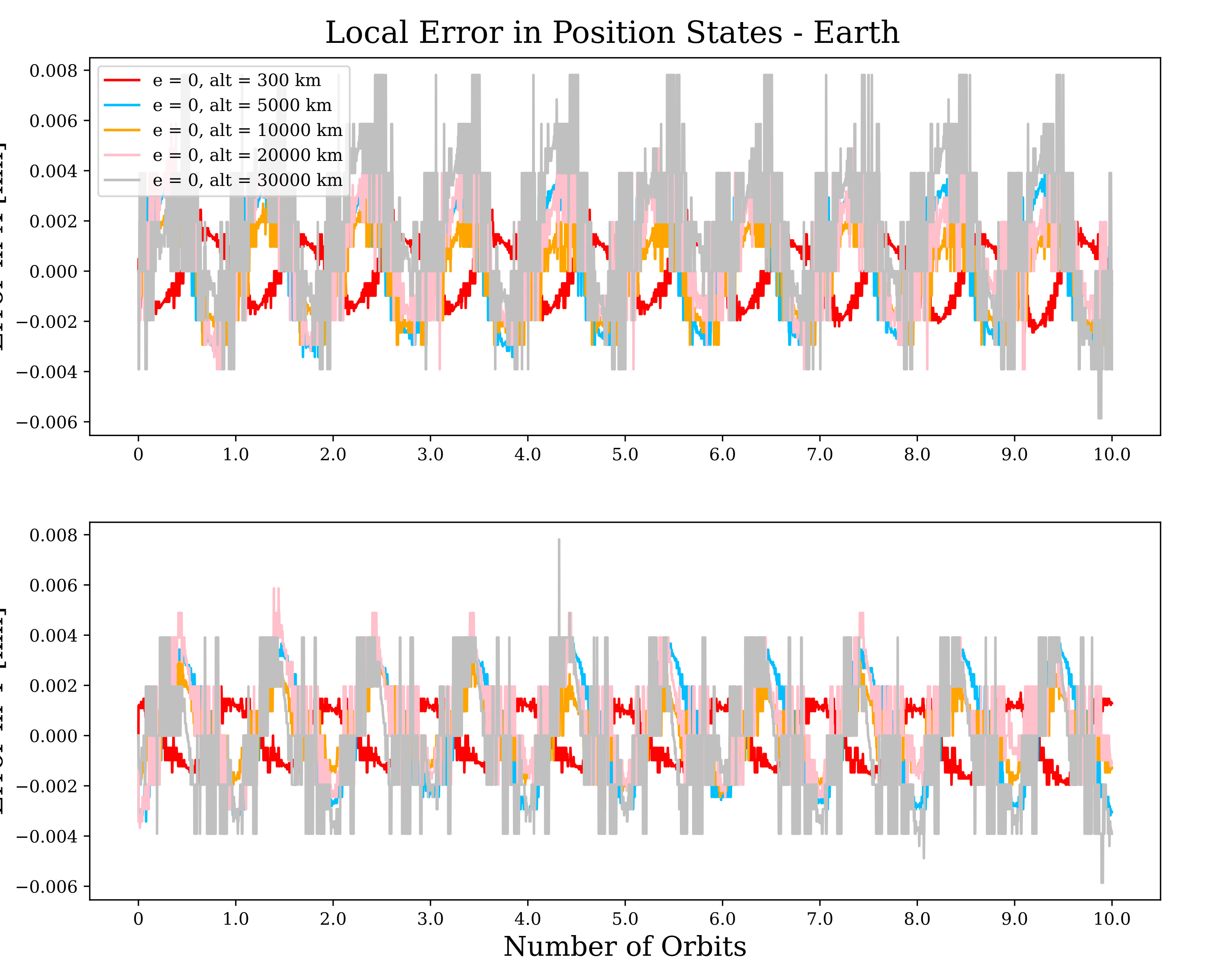}
    \caption{Local error in position states over ten orbital periods around Earth.}
    \label{fig:local_error_10}
\end{figure}

It is clear however, that as the orbit progresses, the prediction error in the position states is somewhat periodic, but not perfectly sinusoidal and appears to grow as the prediction increases. As such, the following simulation demonstrates the prediction of the linear Koopman model over a length of 10 orbital periods, a length of time 10 times greater than any of the training data the model was exposed to. Figure \ref{fig:states_10} displays the four states over the entire simulation duration, where it is evident that the linear Koopman model closely follows the nonlinear dynamics. Again, the global and local errors are presented to thoroughly analyze the ability of the model to closely predict the states of system with Figures \ref{fig:global_error_10} and \ref{fig:local_error_10}. We can see that although the global error slowly grows as the prediction time evolves, the local error is relatively unchanged regardless of the time length as expected. We note that the maximum global positional error is only $1.8\%$ of the orbital radius.

\begin{figure}[H]
    \centering
    \includegraphics[width = 0.85\columnwidth]{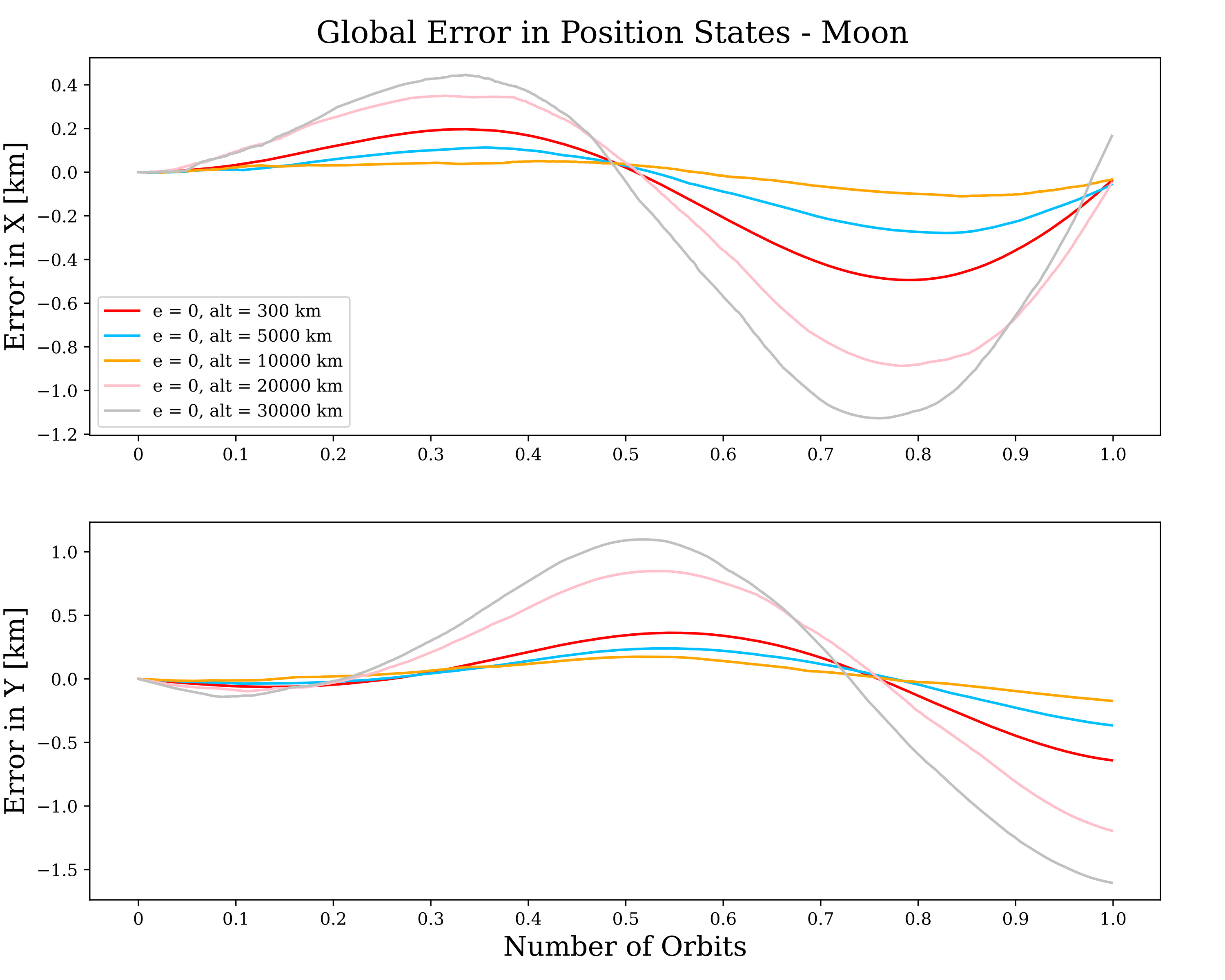}
    \caption{Global error in position states over one orbital period for the Moon.}
    \label{fig:global_error_moon}
\end{figure}

\begin{figure}[H]
    \centering
    \includegraphics[width = 0.85\columnwidth]{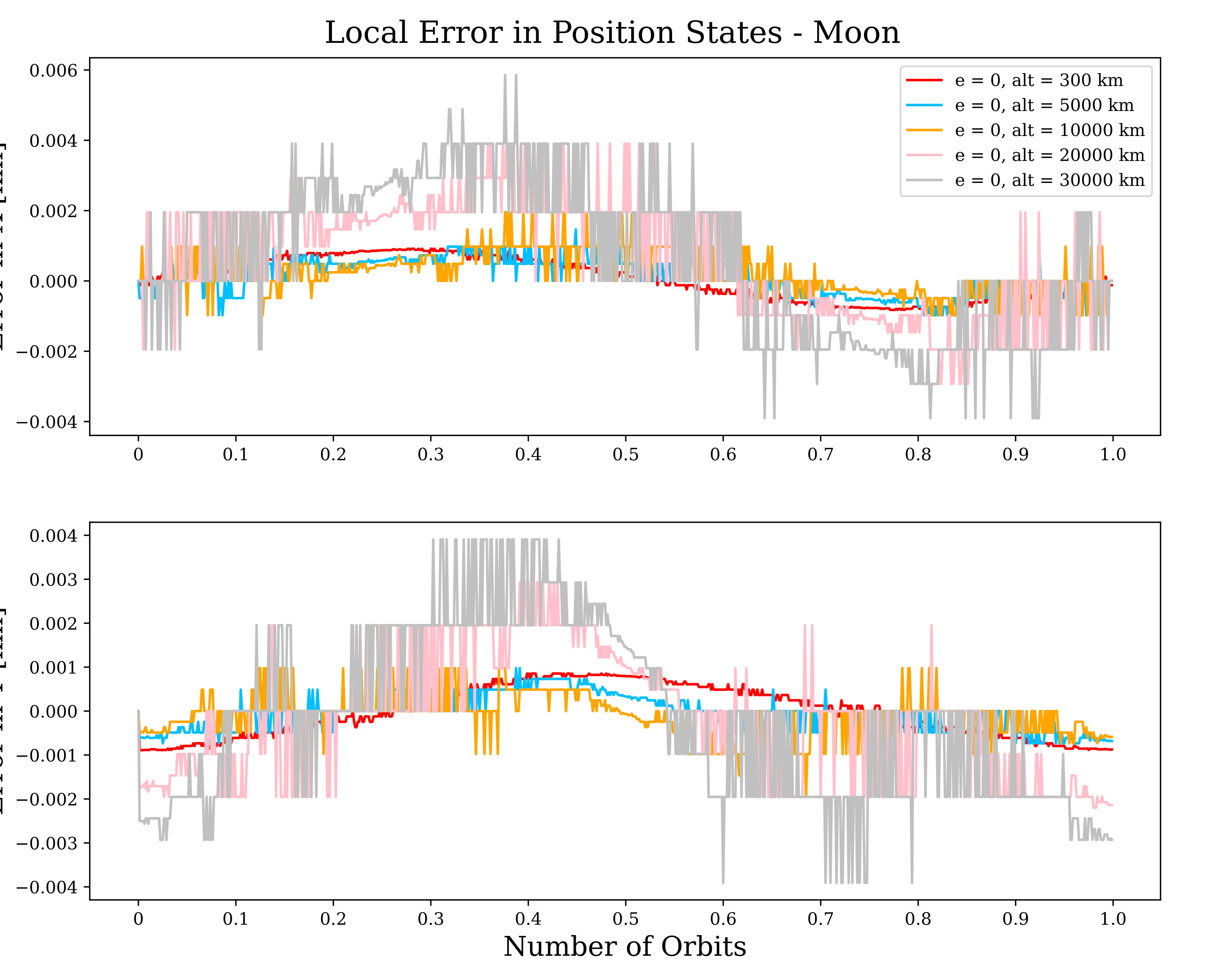}
    \caption{Local error in position states over one orbital period for the Moon.}
    \label{fig:local_error_moon}
\end{figure}

\begin{figure}[H]
    \centering
    \includegraphics[width = 0.85\columnwidth]{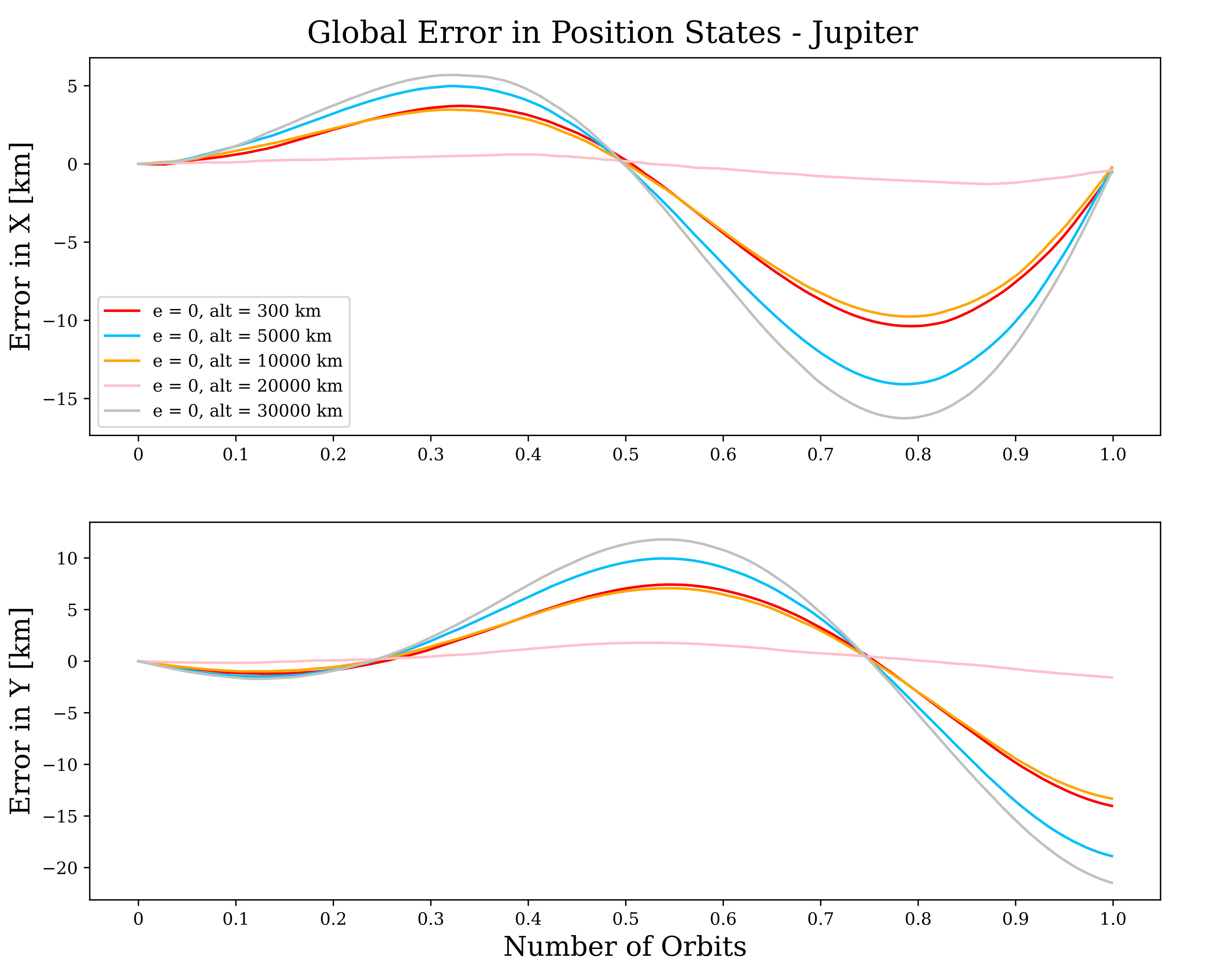}
    \caption{Global error in position states over one orbital period for Jupiter.}
    \label{fig:global_error_jupiter}
\end{figure}

\begin{figure}[H]
    \centering
    \includegraphics[width = 0.85\columnwidth]{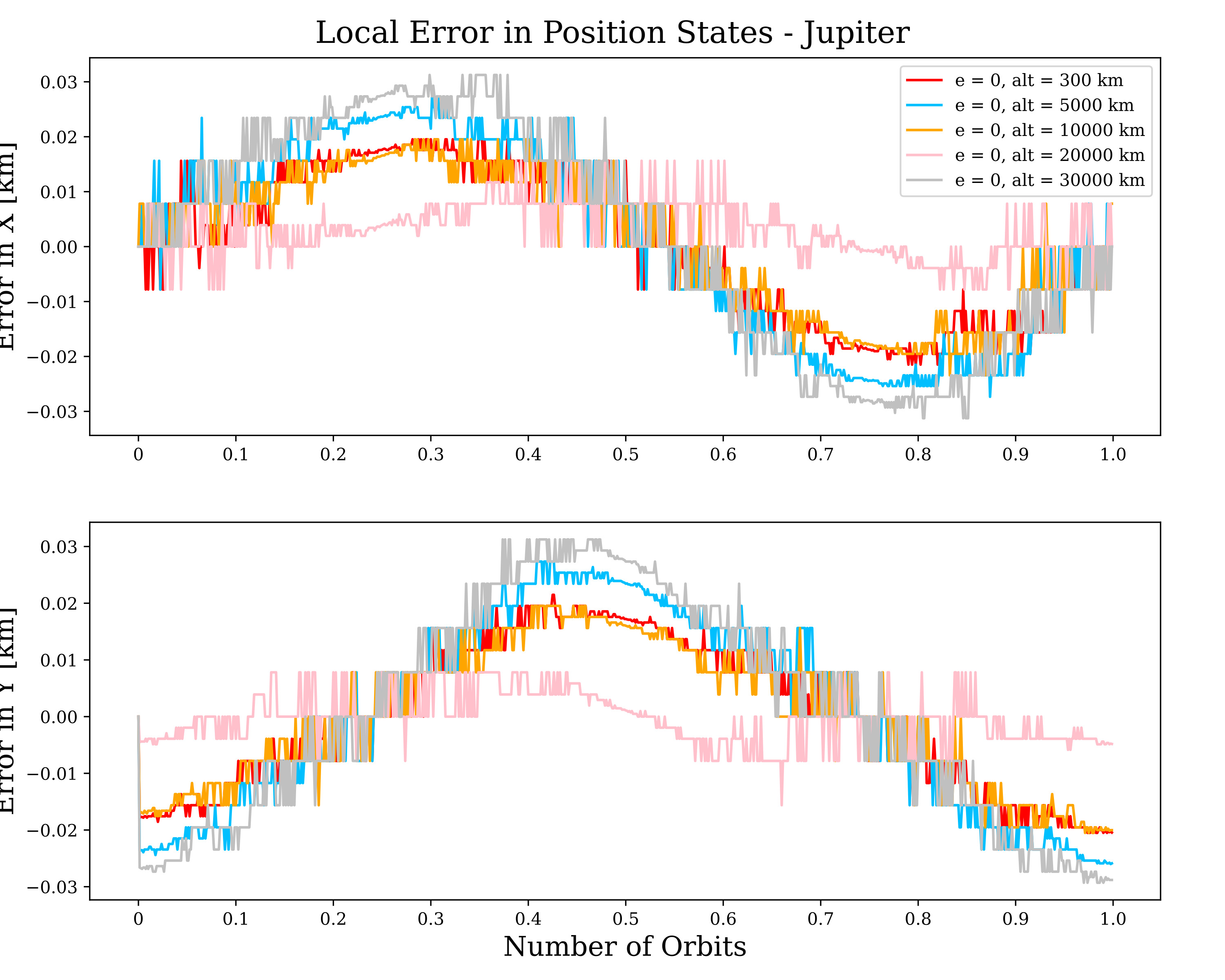}
    \caption{Local error in position states over one orbital period for Jupiter.}
    \label{fig:local_error_jupiter}
\end{figure}

The Koopman operator, although excellent at predicting learned nonlinear dynamics, can often struggle when generalized and implemented in other systems. Here, we show that the model proposed in this work generalizes quite accurately to two other 2BP's of varying sizes. In particular, orbital trajectories around the Moon (for a smaller central body) and Jupiter (for a larger central body) were predicted using the same Koopman operator learned from the dataset based on Earth's orbit. As seen in Figures [\ref{fig:global_error_moon}- \ref{fig:local_error_jupiter}], the global and local errors in both the Moon's and Jupiter's orbits are slightly higher than that of the Earth's, but the prediction still is highly effective. 

\begin{figure}[H]
    \centering
    \includegraphics[width = 0.85\columnwidth]{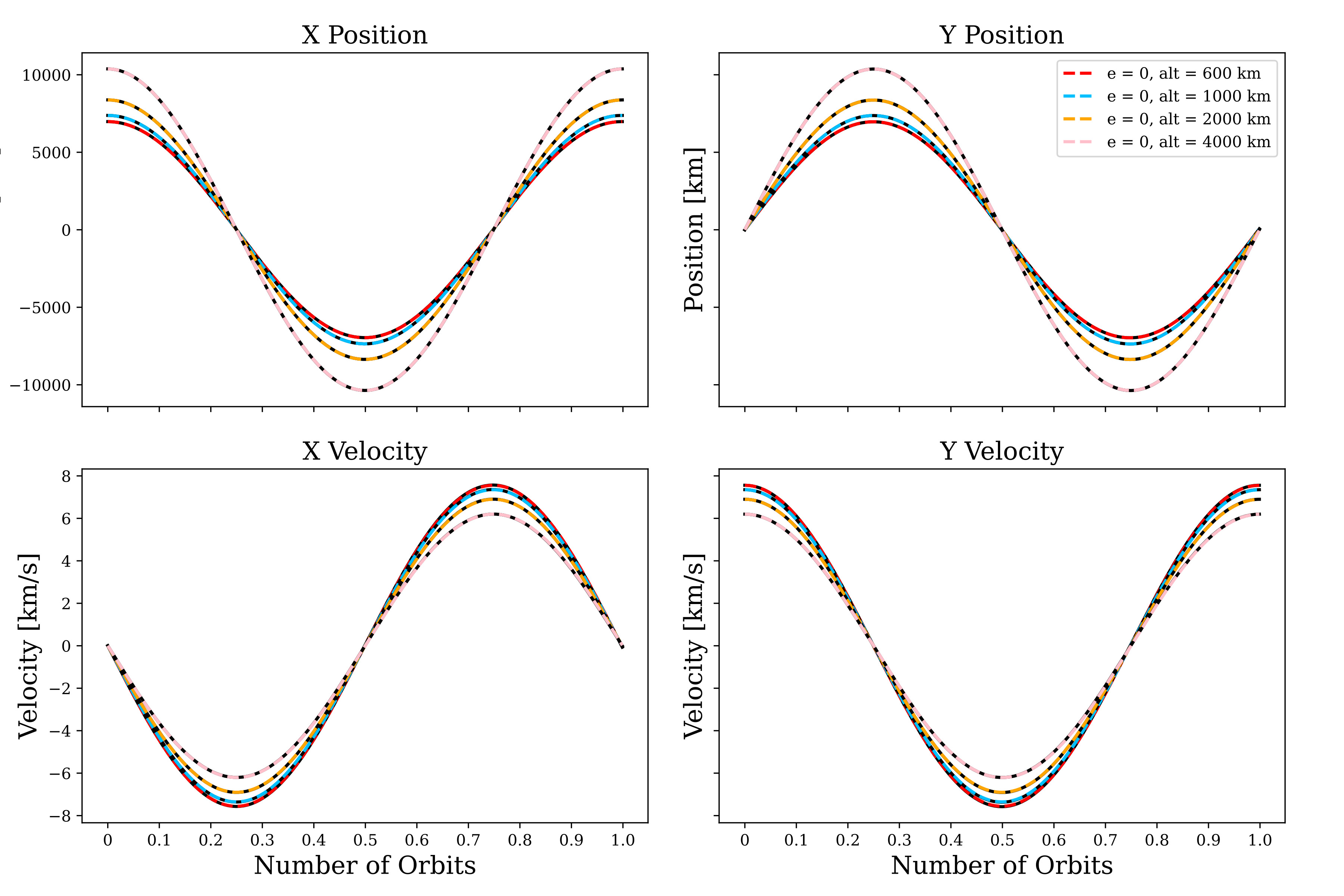}
    \caption{Four position and velocity states of the J2 and SRP perturbed orbits over one orbital period in comparison to the full nonlinear dynamics.}
    \label{fig:states_perturb}
\end{figure}

\begin{figure}[H]
    \centering
    \includegraphics[width = 0.85\columnwidth]{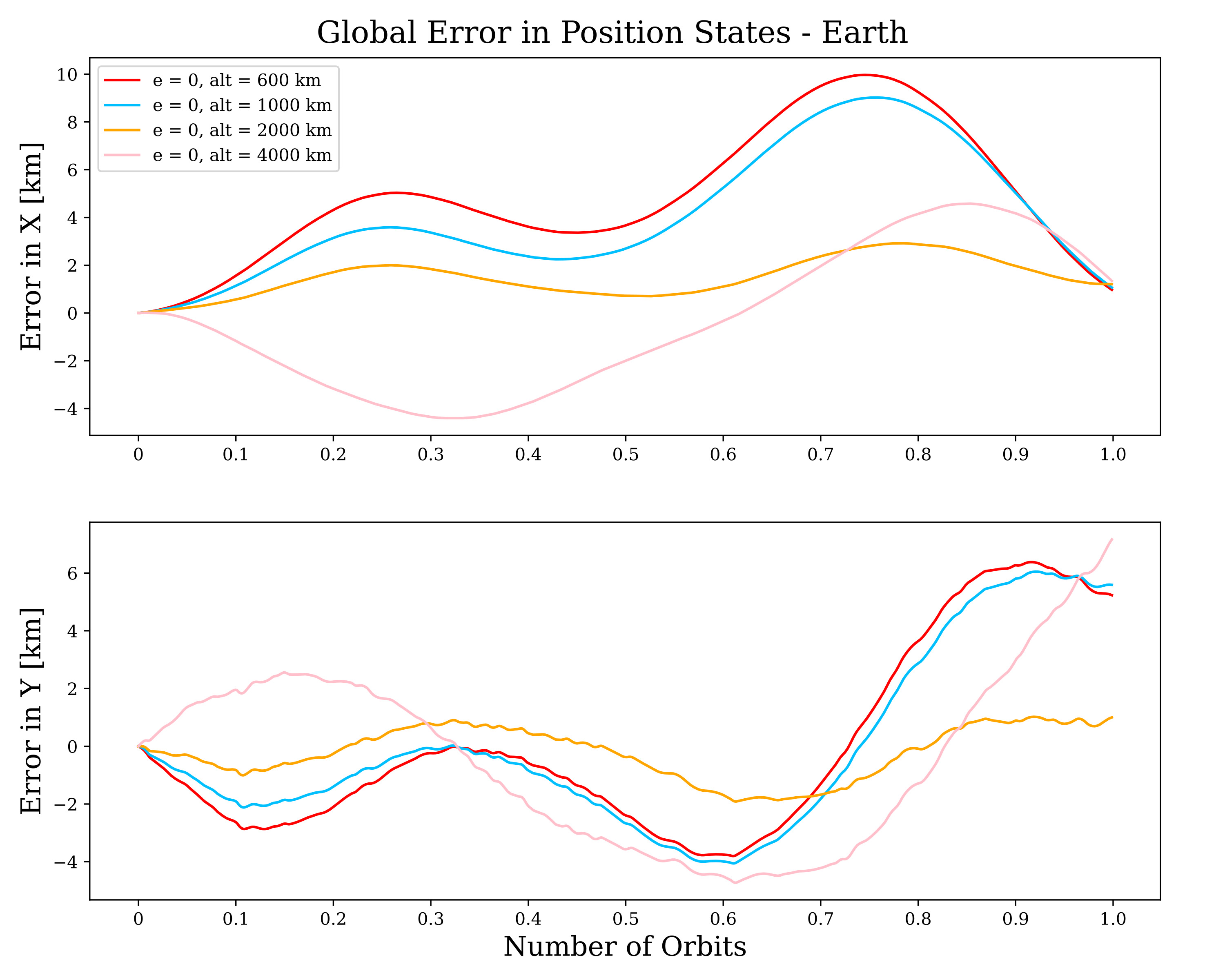}
    \caption{Global error in position over one orbital period of the J2 and SRP perturbed orbit.}
    \label{fig:global_error_perturb}
\end{figure}

\begin{figure}[H]
    \centering
    \includegraphics[width = 0.85\columnwidth]{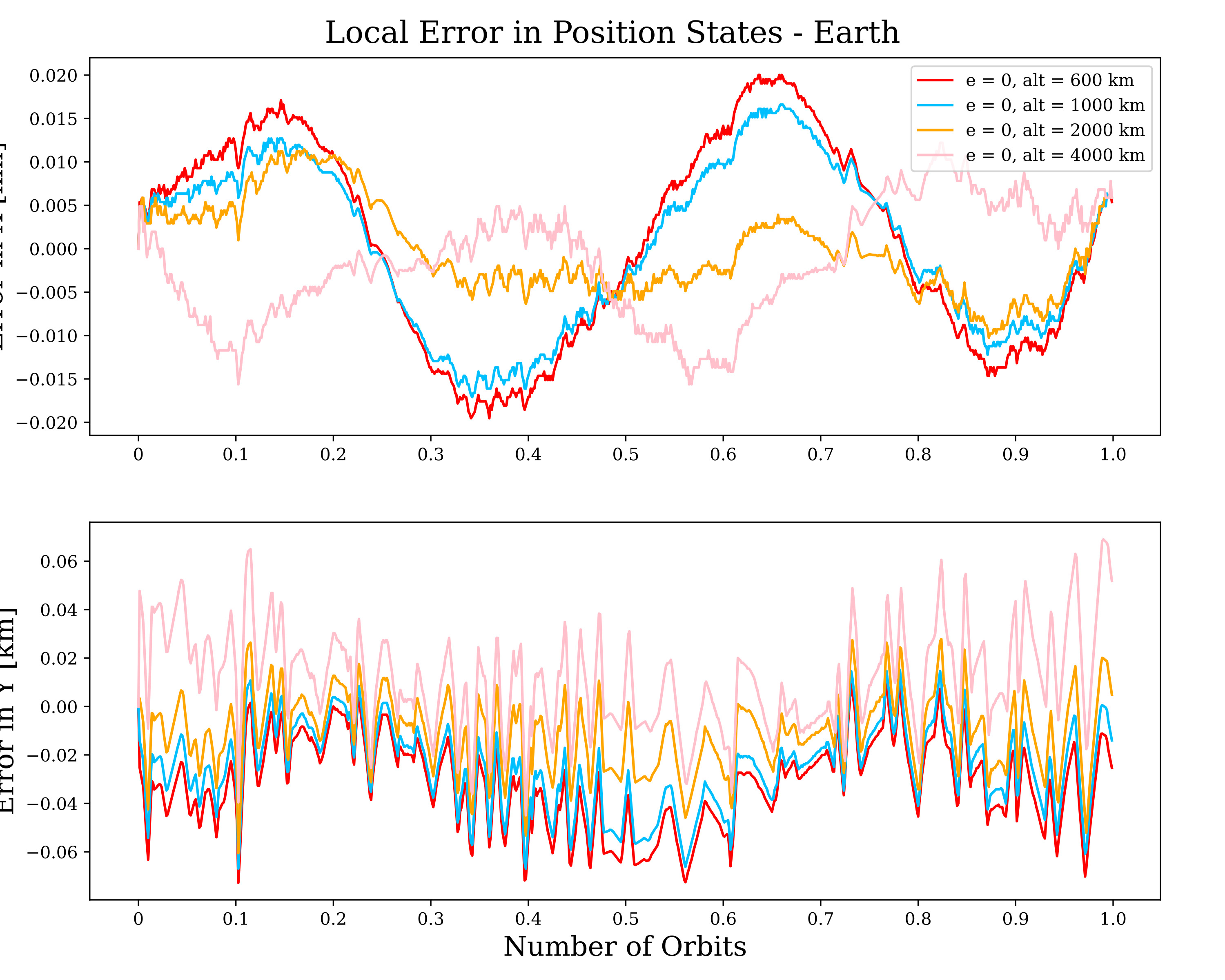}
    \caption{Local error in position over one orbital period of the J2 and SRP perturbed orbit.}
    \label{fig:local_error_perturb}
\end{figure}

In order to truly analyze the generality of our model to capture many of the key aspects of the real 2BP this next simulation presents the scenario where a circular orbit around Earth is perturbed due to solar radiation pressure and the J2 gravity model. Utilizing a training dataset which includes these perturbations, we are able to use our same framework with only a modification in the number of neurons per hidden layer from 25 to 35, to achieve reasonably accurate results. The predicted states of the system show close correspondence to the nonlinear system as shown in Figure \ref{fig:states_perturb}, whilst the global and local errors in Figures \ref{fig:global_error_perturb} and \ref{fig:local_error_perturb} remain within $1.3\%$ of the orbital radius. It can be noted that further tuning and adjustments to the hyperparameters of the network would aid the ability of the model to account for these and other perturbations that are present in realistic simulations. 

\begin{figure}[H]
    \centering
    \includegraphics[width = 0.85\columnwidth]{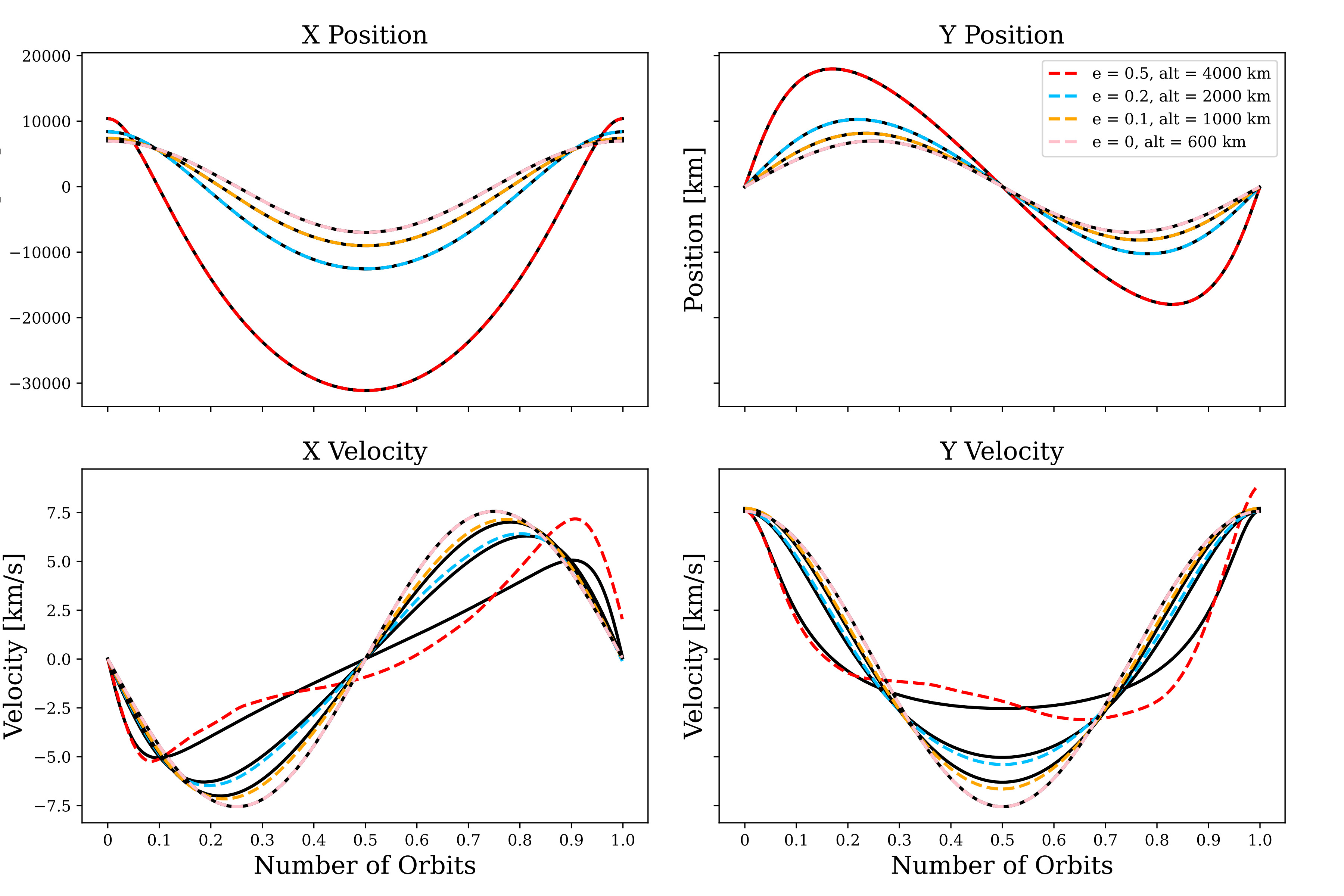}
    \caption{States of the eccentric Earth orbits over one orbital period in comparison to the full nonlinear dynamics.}
    \label{fig:states_ecc}
\end{figure}

\begin{figure}[H]
    \centering
    \includegraphics[width = 0.85\columnwidth]{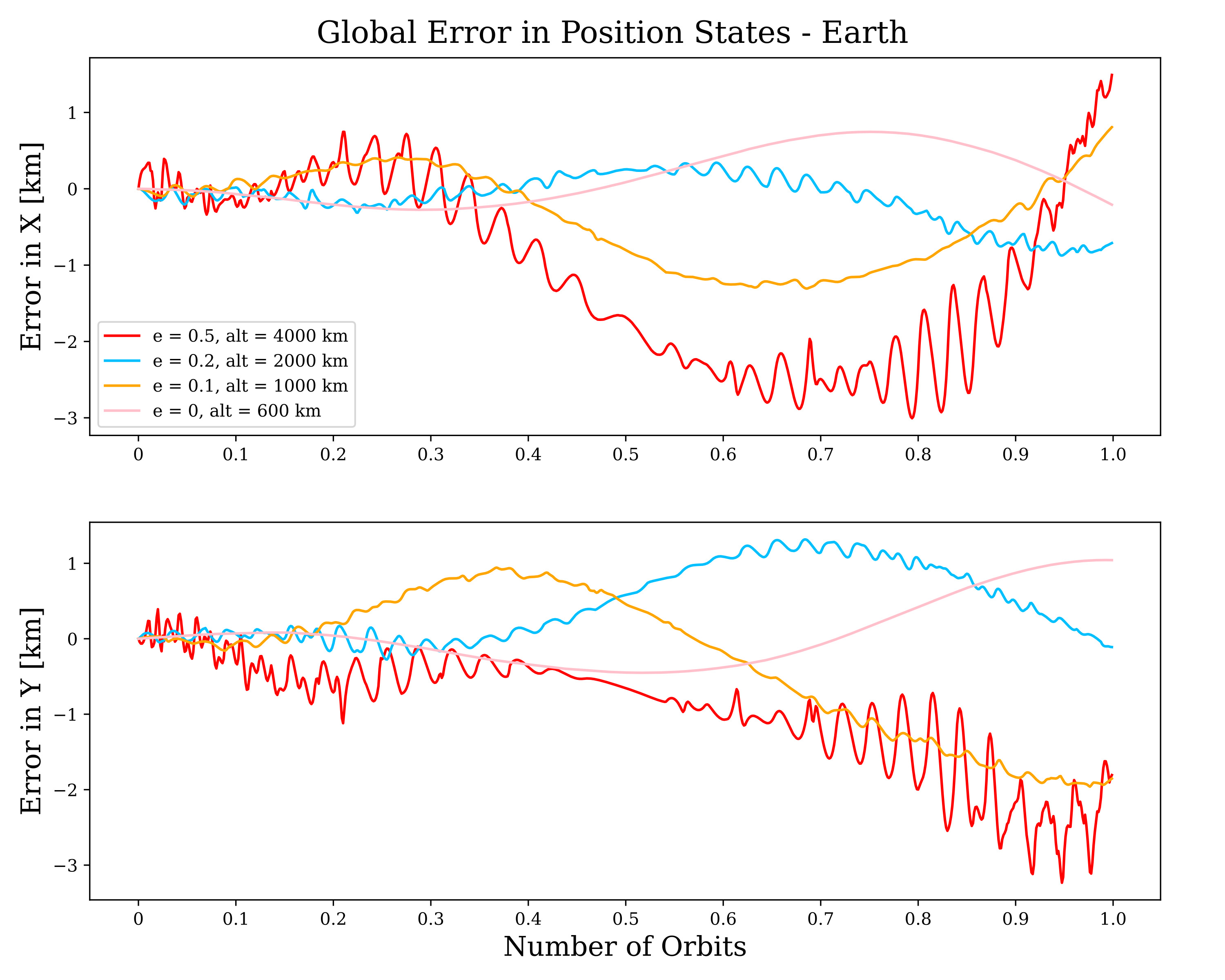}
    \caption{Global error in eccentric position states over one orbital period around Earth.}
    \label{fig:global_error_ecc}
\end{figure}

\begin{figure}[H]
    \centering
    \includegraphics[width = 0.85\columnwidth]{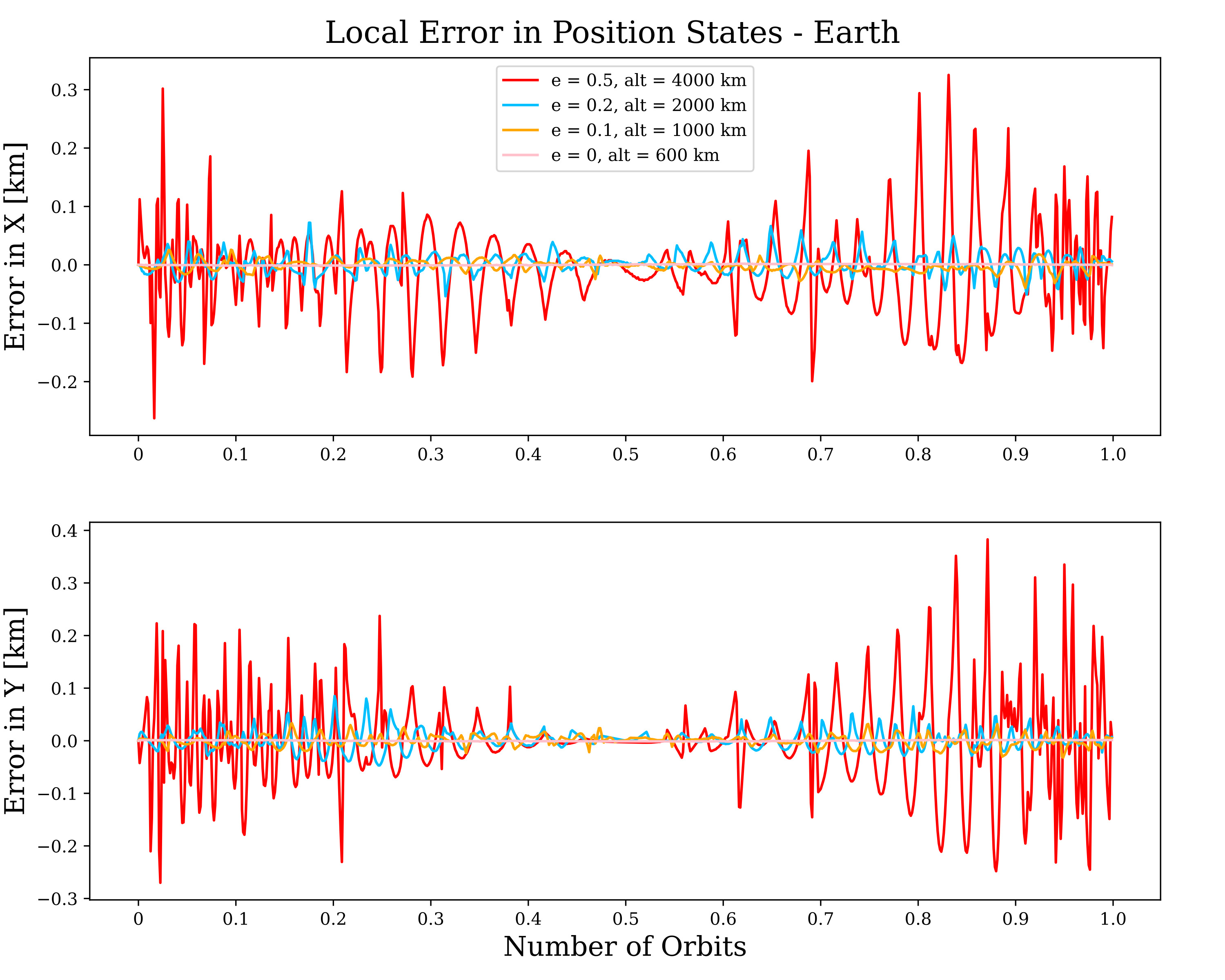}
    \caption{Local error in eccentric position states over one orbital period around Earth.}
    \label{fig:local_error_ecc}
\end{figure}

The next set of simulations come from modeling an eccentric 2BP where the eccentricities of the orbits are 0.1, 0.2, and 0.5. For this simulation, nonlinear orbits with varying eccentricities were included in the training dataset to enable the model to learn this particular system. The states of the various eccentric orbits are presented in Figure \ref{fig:states_ecc}, whilst the global and local errors, as shown previously are given in Figures \ref{fig:global_error_ecc} and \ref{fig:local_error_ecc}. Whilst the errors in position are relatively low and comparable to that of the purely circular or perturbed circular orbits, it is evident from Figure \ref{fig:states_ecc} that for orbits with increasing eccentricities, the velocity prediction begins to degrade under the current conditions. 

\begin{figure}[H]
    \centering
    \includegraphics[width = \columnwidth]{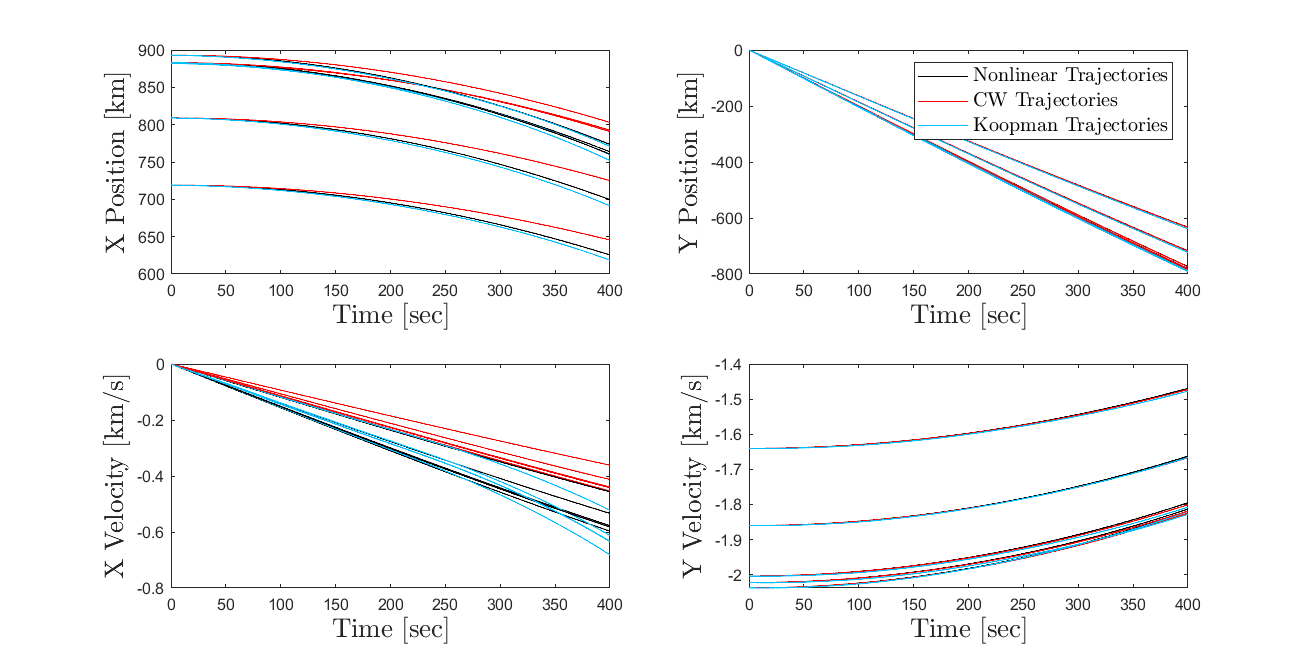}
    \caption{Comparison of the relative states between our Koopman framework, the CW equations and the nonlinear solution.}
    \label{fig:relative_states}
\end{figure}

The Clohessy-Wiltshire (CW) equations, which are a popular form of linearized orbital mechanics deal with the relative motion between a chief and deputy spacecraft. As such, we compare this framework, modified to suit the relative orbital dynamics problem, in our work \cite{Nehma2025-hl}. We invite the reader to read \cite{Nehma2025-hl} for a more comprehensive analysis on how the Koopman operator can be used to relax some of the assumptions needed in the development of the CW equations. We present however, Figure \ref{fig:relative_states} which highlights the ability of our Koopman framework to be quickly applied with minor modifications to this differing scenario. Presented is a comparison between our Koopman model, the CW equation solution and the nonlinear solution for a 200 second Monte Carlo simulation in which the chief spacecraft is in an elliptical orbit and the relative distance between both spacecraft is large, hence violating both assumptions of the CW equations. The true anomaly used in the CW equations, although incorrect, is an average true anomaly for the orbit and allows for a estimated comparison. 

Plots of the error in the position state for both the Moon- and Jupiter-centered orbits allow us to analyze the effect of the size of the central body on the models ability to evolve the dynamics. A larger central body appears to have a greater periodic prediction error as compared to a central body that is smaller than that of the body the training data was derived from. As such, the possibility for the model to be used as a baseline, with additional data from the applied system added online, could enhance the performance of the model. With a short training time of around 120 min, it is not unreasonable to engage in the possibility of online data collection and learning to further enhance the accuracy of the predicted orbit.

\subsubsection{Accuracy Metrics}

\begin{figure}[H]
    \centering
    \includegraphics[width = 0.9\columnwidth]{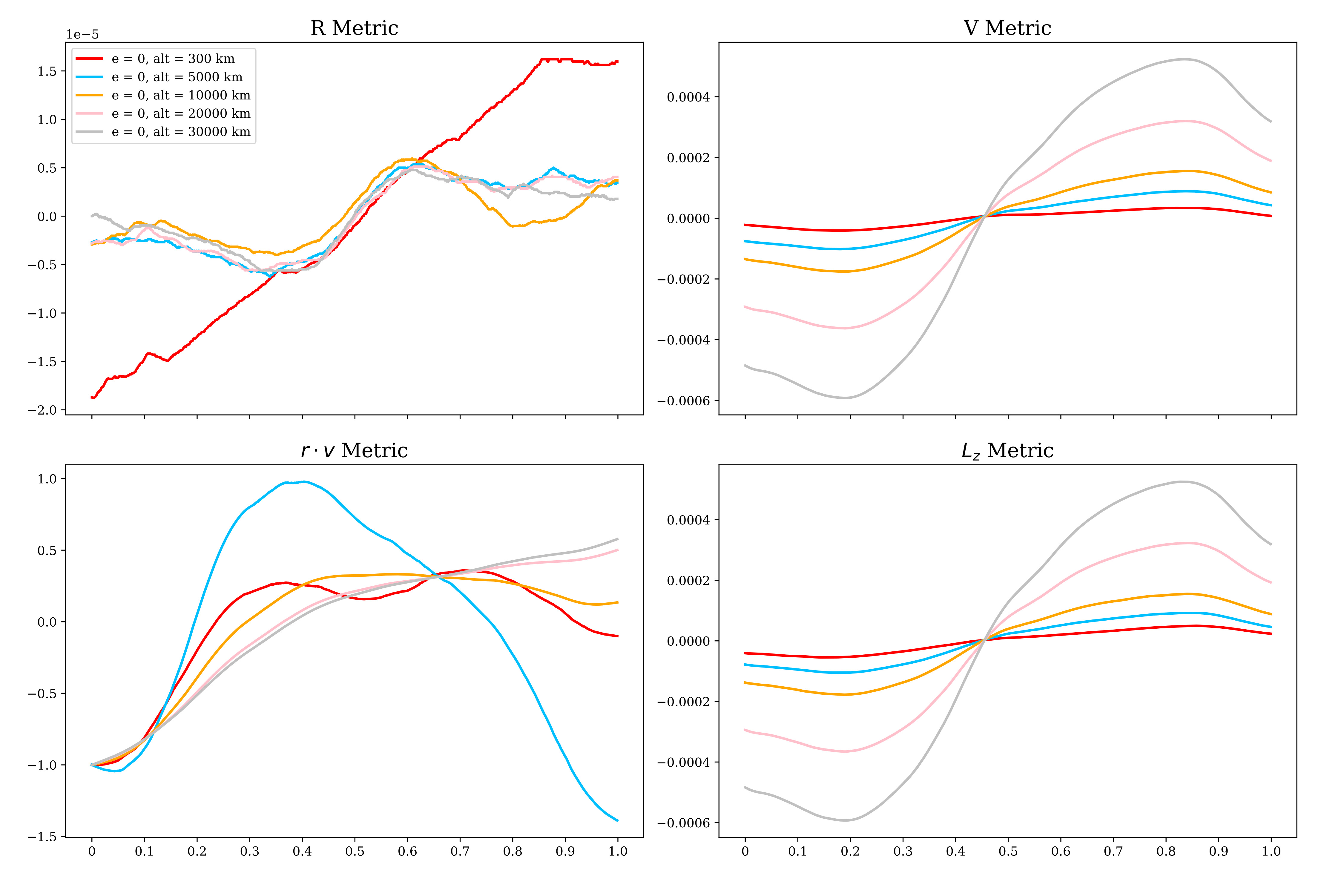}
    \caption{Invariant property metrics for the 2BP linearization around Earth.}
    \label{fig:metric_earth}
\end{figure}

\begin{figure}[H]
    \centering
    \includegraphics[width = 0.85\columnwidth]{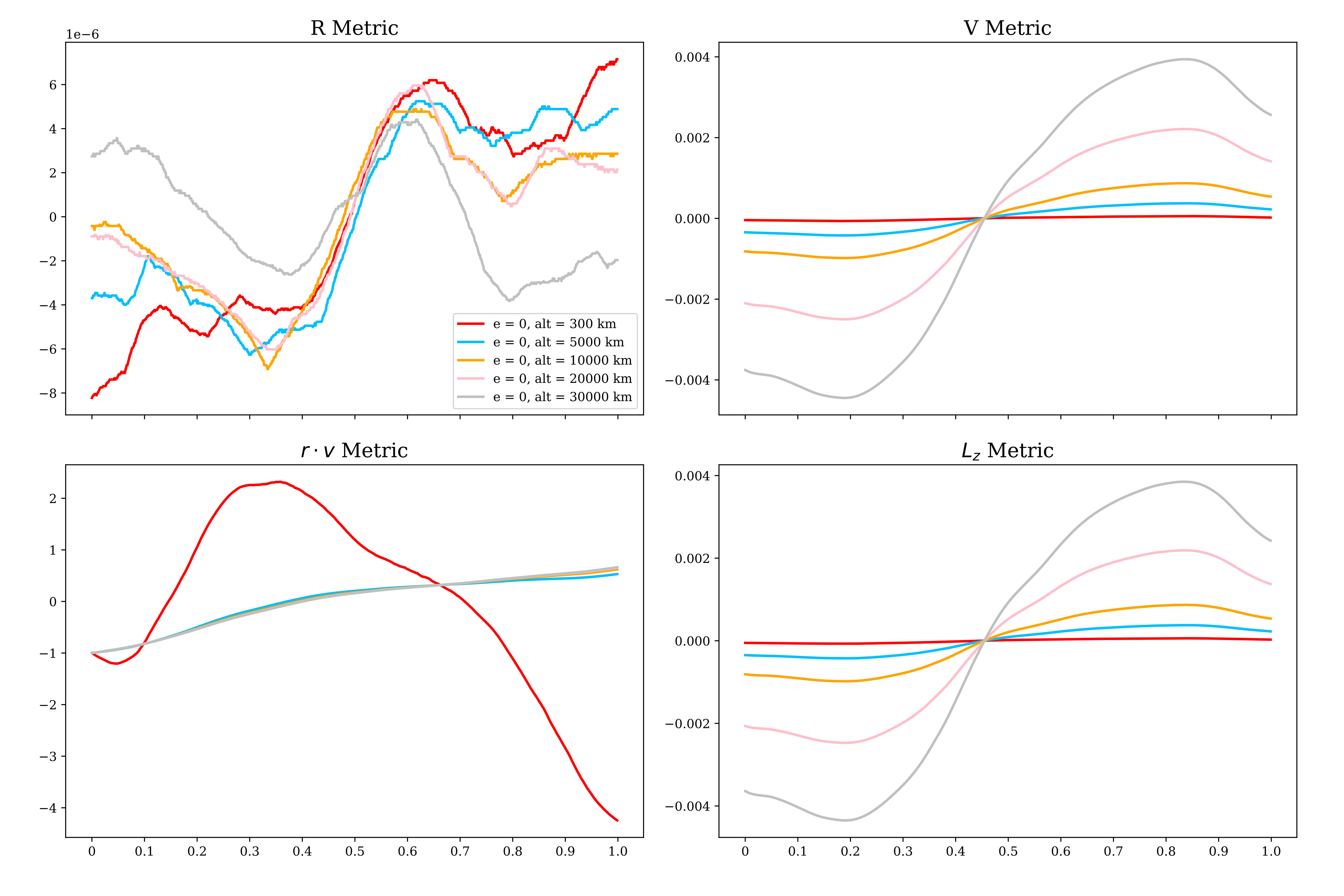}
    \caption{Invariant property metrics for the 2BP linearization around the Moon.}
    \label{fig:metric_moon}
\end{figure}

\begin{figure}[H]
    \centering
    \includegraphics[width = 0.85\columnwidth]{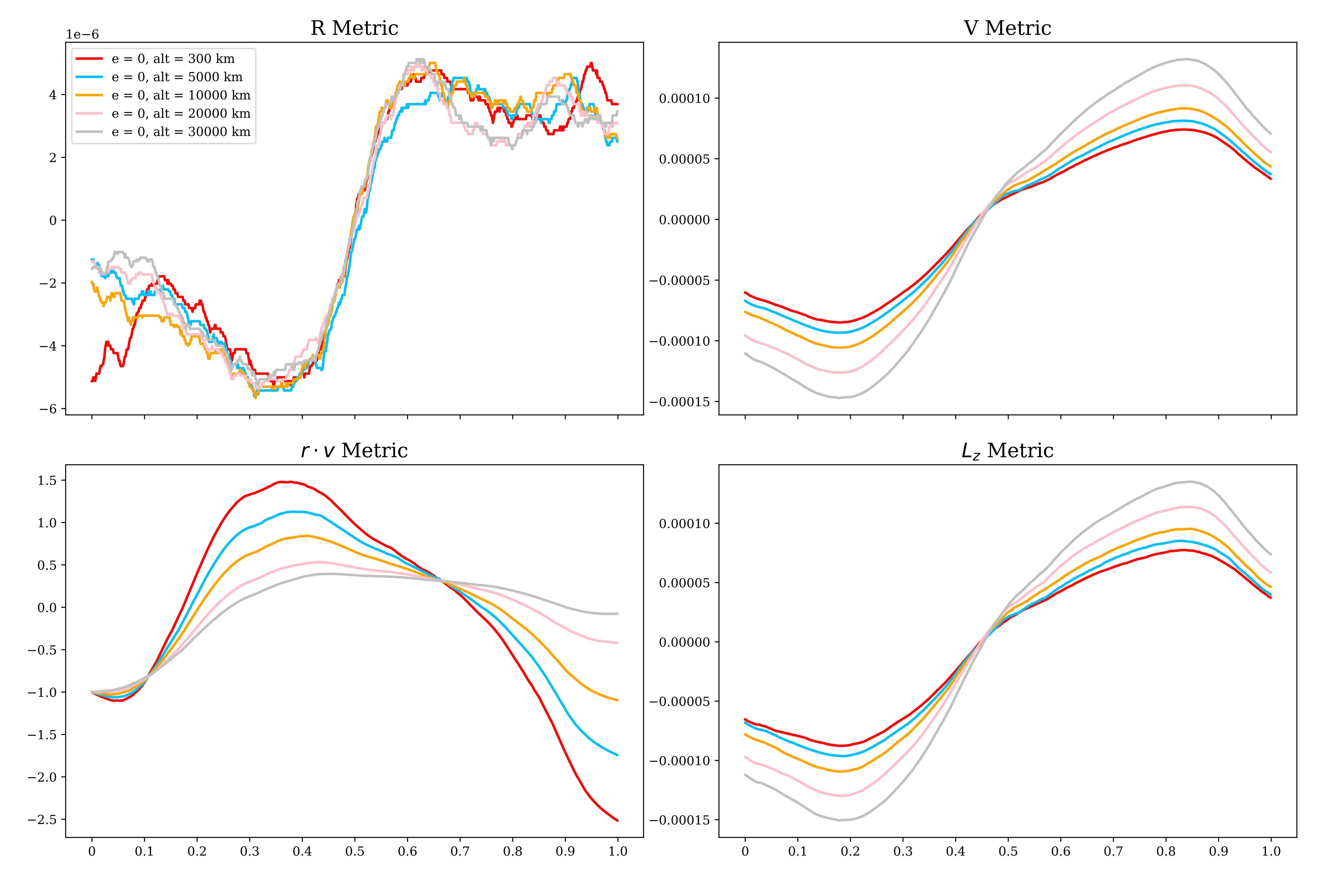}
    \caption{Invariant property metrics for the 2BP linearization around Jupiter.}
    \label{fig:metric_jupiter}
\end{figure}

\begin{figure}[H]
    \centering
    \includegraphics[width = 0.85\columnwidth]{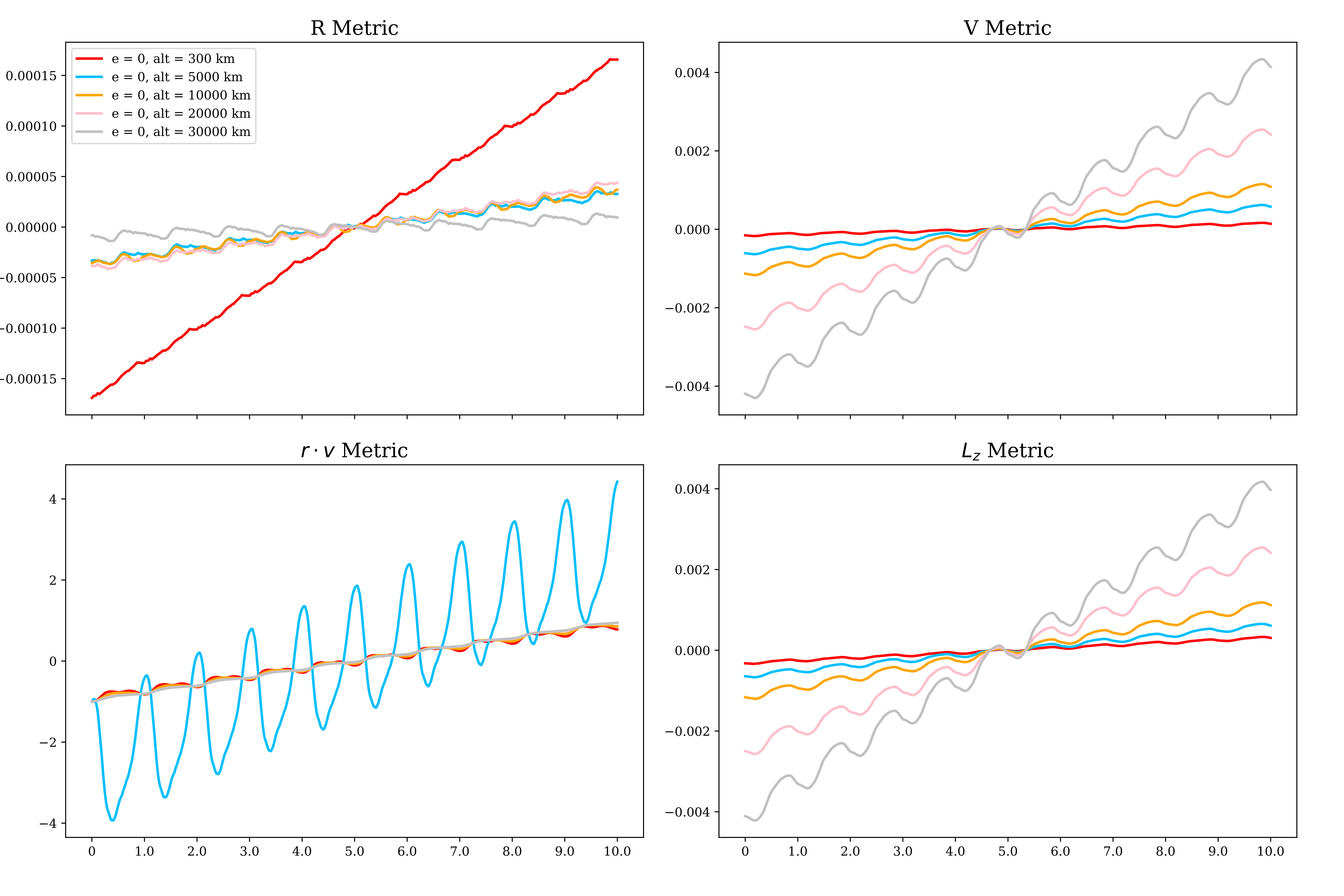}
    \caption{Invariant property metrics for the 2BP linearization over ten orbits.}
    \label{fig:metric_10}
\end{figure}

\begin{figure}[H]
    \centering
    \includegraphics[width = 0.85\columnwidth]{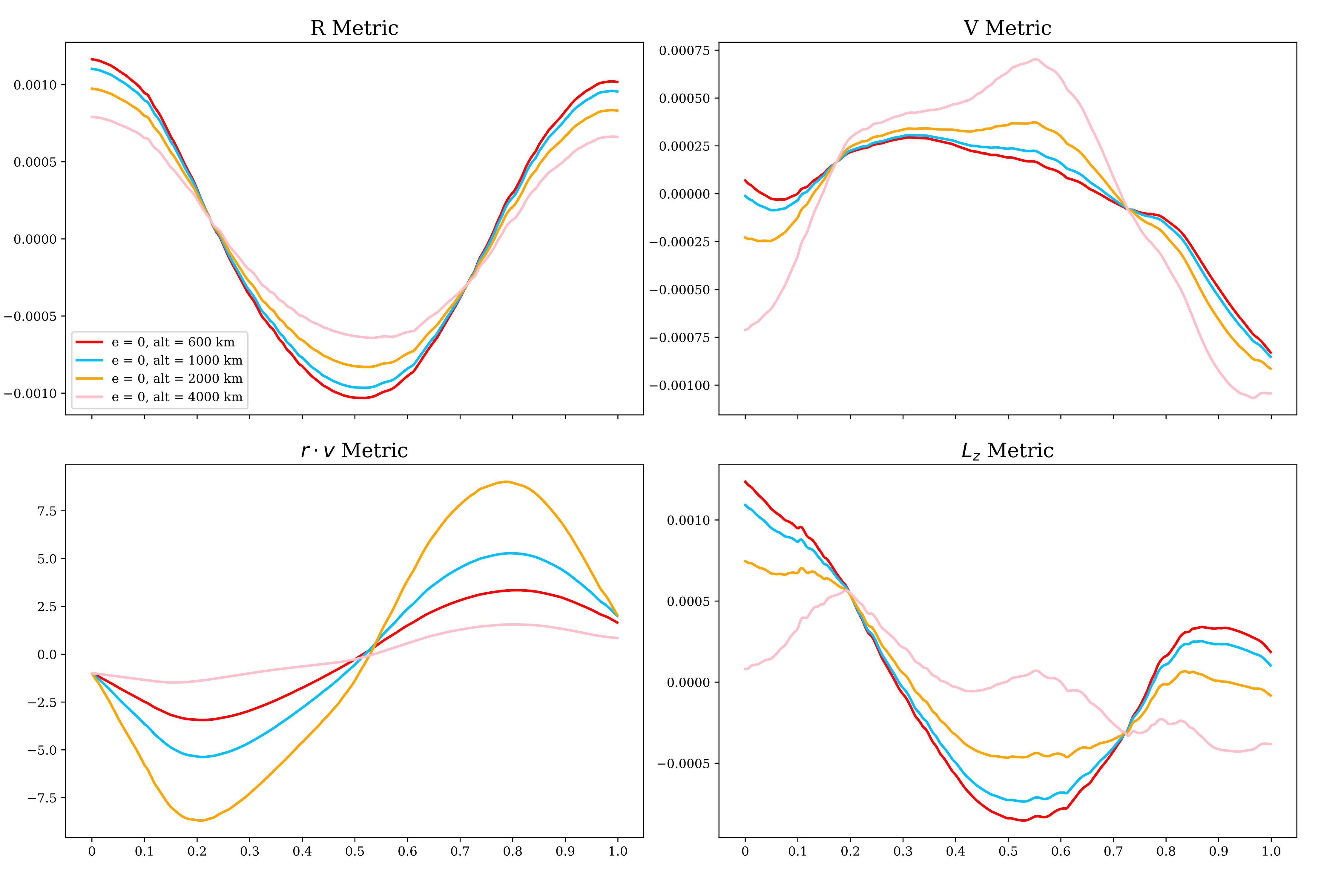}
    \caption{Invariant property metrics for the J2 and SRP perturbed 2BP linearization.}
    \label{fig:metric_perturb}
\end{figure}

We now look at the metrics provided in Section \ref{2BPmetrics}, and how the proposed Koopman operator is able to adhere to the conservative properties aforementioned. We see in Figure \ref{fig:metric_earth} that the values of all four metrics for the model around Earth are substantially low and very close to zero, hence illustrating the absolute error in each metric along the orbit is near zero. Furthermore, when these metrics are checked on orbits around both the Moon and Jupiter, over 10 orbital periods and with the J2 and SRP perturbations, Figures [\ref{fig:metric_moon}- \ref{fig:metric_perturb}], the conservation of the invariant properties is also preserved, as the values are also near zero.

The inclusion of the $r\cdot v$ loss function was specifically included because this metric has the highest degree of error among all other metrics. As seen with the 300 km Moon orbit, 5000 km orbit for 10 periods, it is possible that this metric be inconsistent with the nonlinear dynamics, however it was observed that after the inclusion of the specific loss function, the error in this metric was significantly reduced. As mentioned in Section \ref{2BPmetrics}, it is important to consider the magnitude of these values and whether averaging implies higher errors. These metrics prove the accuracy and reliability of our model in representing the 2BP not only insofar as its specific trained dataset is concerned, but also in its generalized applications to other systems.

\subsubsection{Circular Restricted Three-Body Problem Linearization}

The hyperparameters of the DNN used for all experiments with the CR3BP are outlined in Table \ref{table:2}. The total training time for the DNN used in the CR3BP model was 55 hours on an NVIDIA GeForce RTX 3090 GPU, which is substantially longer than that of the 2BP due to its highly increased lifted space and dataset size. However, it is important to note that this training can occur offline prior to the models use and would not be an issue in the online deployment of the model.

\begin{table}[h]
\begin{center}
\begin{tabular}{ |p{5cm}||p{6cm}| }
 \hline
 \multicolumn{2}{|c|}{\textbf{CR3BP Neural Network Hyperparameters}} \\
 \hline
 \textbf{Hyperparameter} & \textbf{Value} \\
 \hline
 Lifted Space Size   & 100\\
 Hidden Layers &   13\\
 Neurons per hidden layer & 105\\
 Batch Size & 16\\
 Learning Rate    &0.000001\\
 Optimizer &   Adam\\
 Activation Function & SELU\\
 Weight Decay & 0.00001\\
 Epochs & 35000\\
 \(\gamma\) & 2\\
 \(\beta\) & 1\\
 \(\lambda_{L_1}\) & 0.004\\
 \(\lambda_{L_2}\) & 0.001\\
 \hline
  K matrix dimension & 106 x 106\\
 \hline
\end{tabular}
\end{center}
\caption{Hyperparameters for CR3BP Neural Network}
\label{table:2}
\end{table}

\begin{figure}[H]
    \centering
    \includegraphics[width = 0.85\columnwidth]{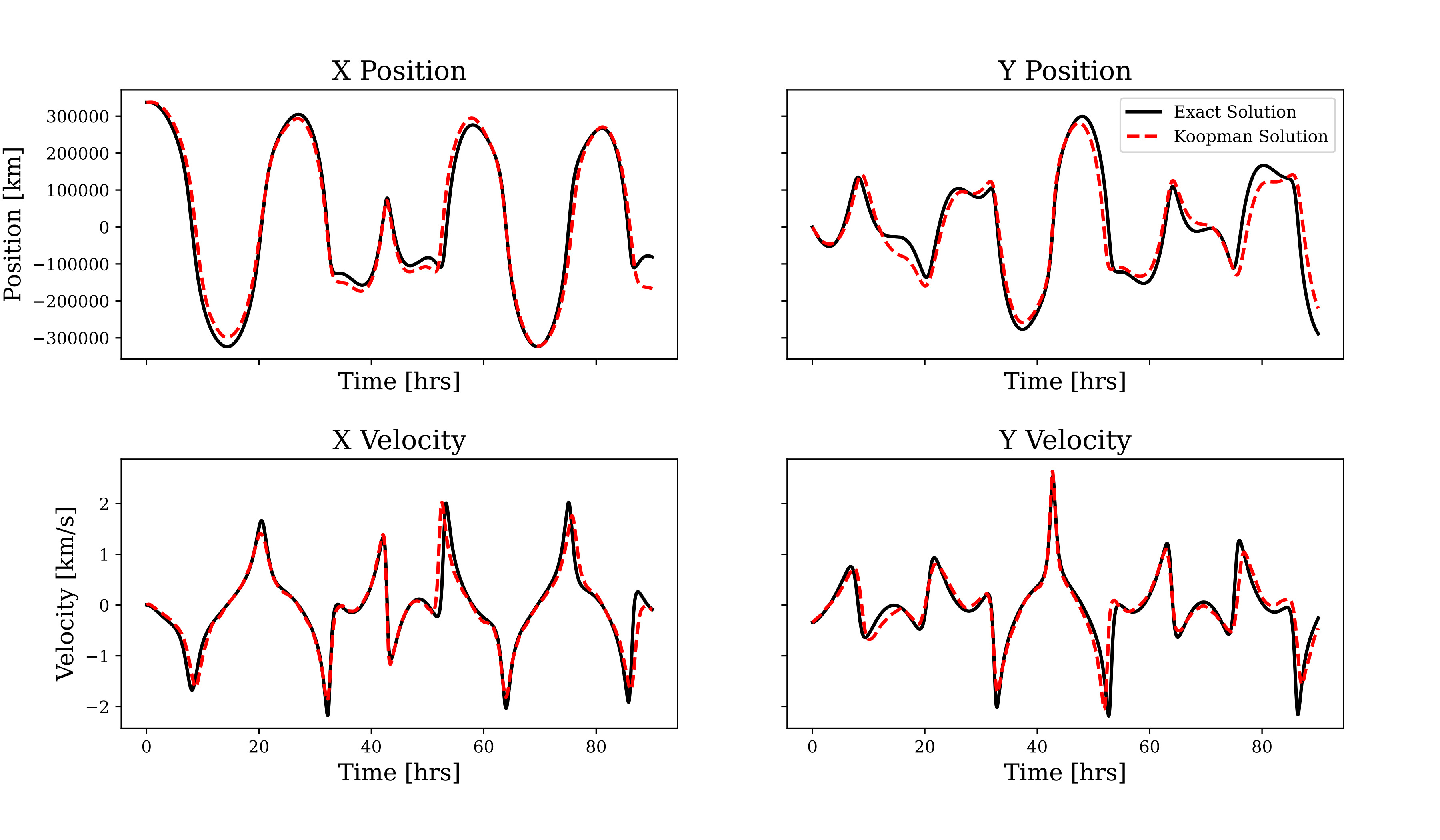}
    \caption{The four states of the CR3BP orbit, with the Koopman model compared to the nonlinear dynamics.}
    \label{fig:CR3BP_states}
\end{figure}

\begin{figure}[H]
    \centering
    \includegraphics[width = \columnwidth]{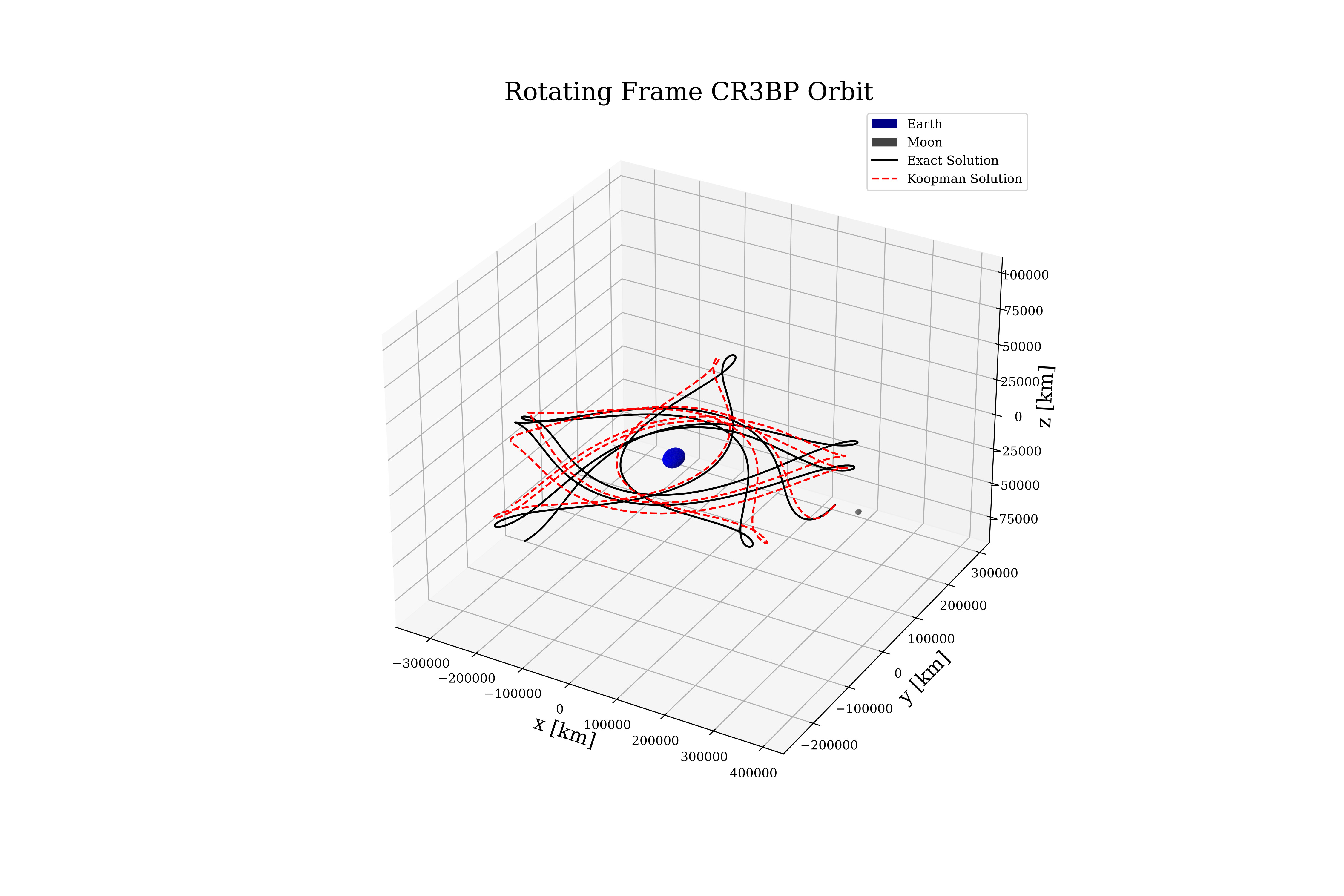}
    \caption{Orbital propagation of the Koopman model and nonlinear dynamics viewed in a 3D plot.}
    \label{fig:CR3BP_3d}
\end{figure}

\begin{figure}[H]
    \centering
    \includegraphics[width = \columnwidth]{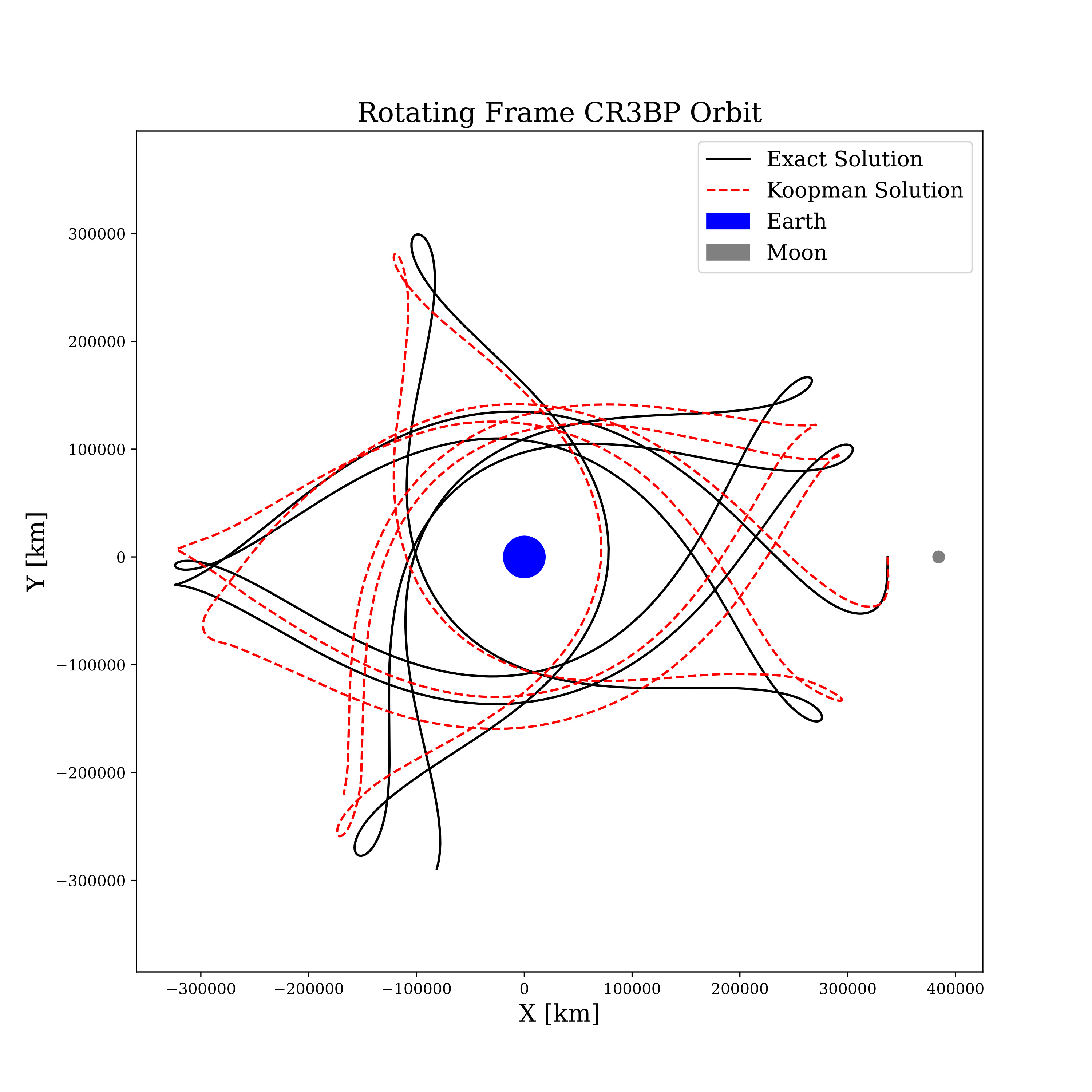}
    \caption{Orbital propagation of the Koopman model and nonlinear dynamics viewed in the 2D orbital plane.}
    \label{fig:CR3BP_2d}
\end{figure}

Figure \ref{fig:CR3BP_3d} shows the linearized evolution of the dynamics that is generated by our Koopman operator. It shows that the DNN is able to learn appropriate observables and lift the dimension of the states such that the K matrix was able to accurately predict the orbit of the CR3BP by rendering it as a linear system. Because the orbit, like the 2BP, is in-plane, the $z$-axis position and velocity are zero, therefore we are able to plot the propagation of the dynamics on a 2D plane to further inspect the accuracy of the propagation, as done in Figure \ref{fig:CR3BP_2d}. Note that in both Figures \ref{fig:CR3BP_3d} and \ref{fig:CR3BP_2d} the scale of the Earth and Moon representations are chosen for ease of viewing, and are not to scale with the propagated orbit. We also show the evolution of each of the 4 states independently in Figure \ref{fig:CR3BP_states}, comparing them to the nonlinear dynamics, to emphasize the ability to closely follow the intricate dynamics of the CR3BP. It shall be noted that although the CR3BP is propagated in the non-dimensionalized units, in these figures the dimensionalized units are restored, and are displayed in the rotating frame. 

The initial condition for the orbit displayed in the aforementioned results was randomly selected from the test data set provided in the generated data set, in order to ensure it is not an IC that the DNN has seen before. The state vector is given as:

\begin{equation*}
    \mathbf{x_0} = \begin{bmatrix}
        0.8673 & 0 & 0 & 0 & -0.2546 & 0
    \end{bmatrix}^{\mkern-1.5mu\mathsf{T}}
\end{equation*}

\subsubsection{Accuracy Metrics}

\begin{figure}[hbt!]
    \centering
    \includegraphics[width = \columnwidth]{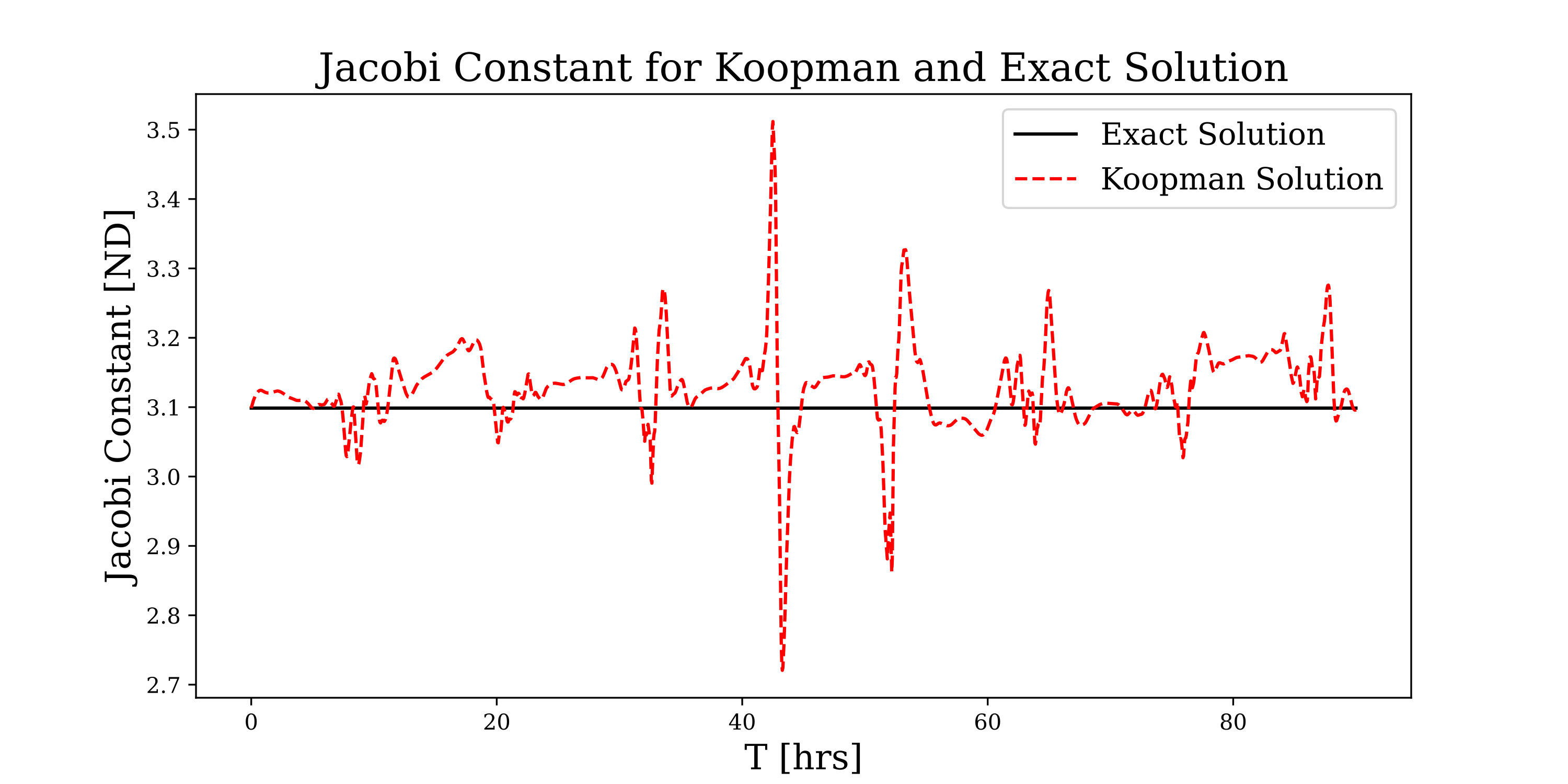}
    \caption{The Jacobi Constant at every time instance for both the nonlinear and linear dynamics.}
    \label{fig:jacobi_constant}
\end{figure}

As mentioned in Section \ref{jacobi}, the metric to determine the accuracy of our linear model is the Jacobi Constant. In order to display this metric, it has been calculated for both the nonlinear and linear models and then plotted against each other. The higher the accuracy of the linear model, the closer the linear Jacobi constant will follow that of the nonlinear dynamics. Figure \ref{fig:jacobi_constant} compares the constant of our model against that of the true, nonlinear dynamics, exhibiting a fair degree of accuracy in light of the complexity of the CR3BP.

\subsection{Training Loss}

\begin{figure}[H]
    \begin{subfigure}{0.5\textwidth}
        \centering
        \includegraphics[width = \linewidth]{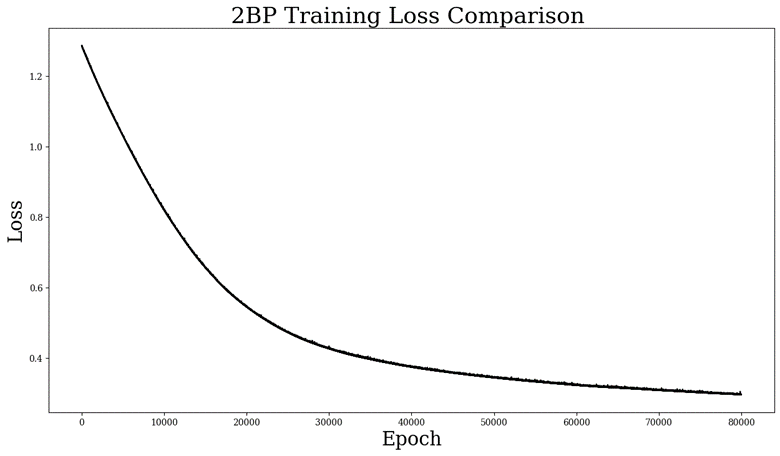}
        \caption{Neural Network training loss across all training epochs for the 2BP.}
        \label{fig:2BP_loss}
    \end{subfigure}
    \begin{subfigure}{0.5\textwidth}
        \centering
        \includegraphics[width = \linewidth]{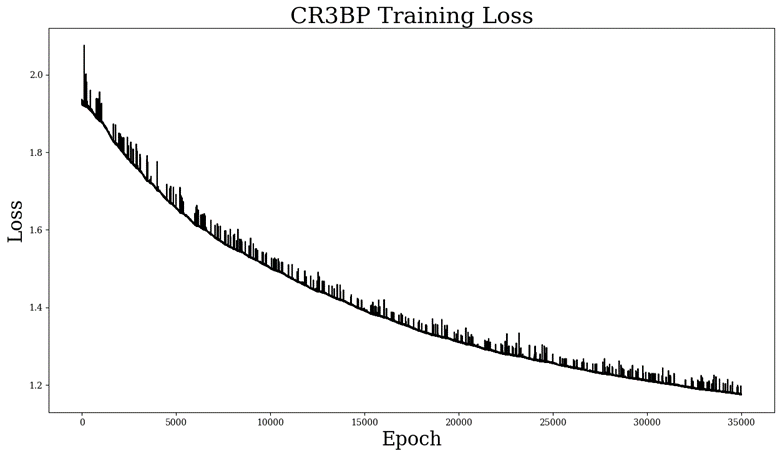}
        \caption{Neural Network training loss across all training epochs for the CR3BP.}
        \label{fig:CR3BP_loss}
\end{subfigure}
\end{figure}

In this section we analyze the training loss for both models. We can see that from both Figures \ref{fig:2BP_loss} and \ref{fig:CR3BP_loss}, the loss function exponentially decays as the number of training epochs increases. This is very typical behavior of a NN model, and shows that the model is adequately trained for each prediction due to its steady state value reached. With more training, tuning of the hyperparameters and loss function it could be reduced even further. 

\subsection{Discussion}

The propagation of the 2BP in the Earth system is very accurate, with minimal global and local error in both the position states and metrics, even for IC's that were outside of the trained region. The largest absolute error in all of the trialed IC's in the Earth system was 2 km, which for that orbit translates to an error percentage of 0.018\% (5000 km orbit). This percentage is calculated as the ratio of the largest error over the mean radius of the orbit. A longer time evolution for the prediction was considered in the next simulation, where 10 orbital periods where prediction from each initial condition. The maximum error percentage experienced in this simulation was $0.18\%$ (5000 km orbit), which although higher than the single orbit, is still relatively low. We notice however that this error grows steadily as the length of the prediction increases, so methods of state estimation in a complete guidance navigation and control loop with this LTI Koopman model would likely help alleviate this growing error by 'resetting' the initial condition. 

By utilizing the LTI Koopman operator learned on the Earth system on the Moon and Jupiter systems, we see that there is a capability for the model to generalize to other dynamical systems, however, the model is sensitive to variations in parameters in the nonlinear dynamics. The error in the predictions varied in comparison to the Earth model because of the dissimilar mass of the central body, ultimately changing its gravitational parameter. The largest error percentage in the Moon system was \(\sim0.0033\%\) (30,000 km orbit) whilst the largest error in the Jupiter system was \(\sim0.015\%\) (30,000 km orbit). Although the errors are still very small, if such a system was to be used in a controls application or a trajectory tracking situation, these errors would have to be reduced further, or have the control system correct it. 

We next demonstrate the capability of the model to handle nonlinear and non-periodic disturbances to the nonlinear dynamics in the form of the J2 gravity perturbation and the solar radiation pressure force. We notice adequate agreeance between the exact nonlinear dynamics and the LTI Koopman model, with a maximum error percentage of $0.14\%$ (300 km orbit). The ability of the model to generalize to eccentric orbits showed that this was where the model struggled with the most. The larger the eccentricity of the orbit, the worse the prediction became, with an eccentricity of 0.5 displaying noticeable inconsistencies between the linear and nonlinear systems. The positional error percentage for these orbits however had a maximum value of $0.04\%$ (300 km orbit). With further concentration in terms of training data, loss function manipulation and hyperparameter tuning it would be possible to tune the model to provide higher degrees of accuracy for a specific application that requires eccentric or perturbed orbits. 

The analysis of the accuracy of the 2BP Koopman model with the metrics in Section \ref{2BPmetrics} shows that our model not only is able to predict the current trajectory of the orbit, but also is able to conserve the physical properties that are evident in the nonlinear dynamics. The learned linear model is even able to conserve these invariants for systems it has never seen before or been trained on, emphasizing the remarkable capabilities of the Koopman operator to effectively linearize a nonlinear system, whilst remaining physically and mathematically valid. Deviating from the purely circular orbits, it is evident that the model struggles with preserving the $r \cdot v$ metric for some of the more complex models simulated. The 300 km orbit around the Moon, 5000 km orbit for ten periods, and a majority of the perturbed circular orbits experienced undesirable values in this metric. Our implementation of the metric-specific loss function reduced the magnitude of the disparity by an order of two, showing it is possible to reduce this further, however conclusive investigation is needed.

The analysis of the Jacobi Constant for the CR3BP model highlights its ability to conserve the physical properties related with its dynamics. Although the accuracy of the CR3BP LTI model is not as high as the 2BP model, it clearly illustrates the ability to capture key important features, with the relative error in the Jacobi Constant being \(\lesssim 12.7\%\). Both the 2BP and CR3BP cases show that the model is accurately learning the dynamics in the linearization and not simply learning shapes or curves. 

Although the 2BP shows excellent results and accuracy in the linear model, the CR3BP apparently still exhibits a higher relative error in its prediction that could be subsequently improved on. The model clearly learns the dynamics of the nonlinear system, but further fine tuning in the loss function, training, hyperparameters or data would be necessary to improve the results to the point that it may be used in a real life problem. 

It is important to note that the linear models presented in this work are LTI, and hence transform the complex nonlinear dynamics into the simple, well-understood form that is the LTI system. Hence, it is possible to employ a wide variety of techniques such as proving stability, observability and controllability that are commonly used by engineers in guidance, navigation, and control.

\section{CONCLUSION}

The advancements in this paper are significant because, to our knowledge, this is the first time that circular, eccentric and perturbed 2BP and the CR3BP, around the L1 Lagrange point, have been globally linearized. Our proposed model, which is also capable of generalization to a wide array of celestial systems, demonstrates a fast and accurate method for developing a linear Koopman operator representation of the 2BP and CR3BP. We establish the accuracy of both models through analysis of the invariant properties of each nonlinear system, and show that each of the invariants hold for our linearized model. 

The LTI models discovered in this paper offer a novel opportunity for several advancements in the control and estimation of two- and three-body problems. The models have diverse applications that can replace a number of older methods such as the CW equations, Taylor series linearization, Kalman filtering and system ID. 

\clearpage

\bibliographystyle{elsarticle-num}
\bibliography{references}

\end{document}